\documentclass[a4paper,fleqn,useAMS,usenatbib]{mnras}
\usepackage[T1]{fontenc}
\usepackage{ae,aecompl}

\usepackage{aas_macros}
\usepackage{graphicx}
\usepackage{amssymb}
\usepackage{amsmath}
\usepackage{widetext}
\usepackage{float}

\usepackage{color}

\title[Central halo assembly and galaxy angular momentum]{The link between the assembly of the inner dark matter halo and the angular momentum evolution of galaxies in the EAGLE simulation}
\author[J. Zavala et al.]{\parbox{17.5cm}{Jes\'us
    Zavala$^{1}$\thanks{Marie Curie Fellow; e-mail: jzavala@dark-cosmology.dk}, Carlos S. Frenk$^{2}$, Richard Bower$^{2}$, Joop Schaye$^{3}$, Tom Theuns$^{2}$, Robert A. Crain$^{4}$, James W. Trayford$^2$, Matthieu Schaller$^{2}$ \& Michelle Furlong$^{2}$\vspace{0.3cm}}\\ 
$^{1}$Dark Cosmology Centre, Niels Bohr Institute, University of Copenhagen, Juliane Maries Vej 30, 2100 Copenhagen, Denmark\\
$^{2}$Institute for Computational Cosmology, Department of Physics, University of Durham, South Road, Durham DH1 3LE, UK\\
$^{3}$Leiden Observatory, Leiden University, PO Box 9513, NL-2300 RA Leiden, the Netherlands\\
$^{4}$Astrophysics Research Institute, Liverpool John Moores University, IC2, 146 Brownlow Hill, Liverpool, L3 5RF, UK}

\date{Accepted XXX. Received YYY; in original form ZZZ}

\pubyear{2015}

\begin{document}
\label{firstpage}
\pagerange{\pageref{firstpage}--\pageref{lastpage}}
\maketitle






\begin{abstract}
We explore the co-evolution of the specific angular momentum of dark matter haloes and the cold baryons 
that comprise the galaxies within. We study over two thousand central galaxies within the reference cosmological hydrodynamical simulation of the ``Evolution and Assembly of GaLaxies and their Environments" (EAGLE) project. 
We employ a methodology within which the evolutionary history of a system is specified by the time-evolving properties of the Lagrangian particles that define it at $z=0$.
We find a strong correlation between the evolution of the specific angular momentum of today's stars (cold gas) and that of the inner (whole) dark matter halo they are associated with. This link is particularly strong for the stars formed before the epoch of maximum expansion and subsequent collapse of the central dark matter halo (turnaround). Spheroids are typically assembled primarily from stars formed prior to turnaround, and are therefore destined to suffer a net loss of angular momentum associated with the strong merging activity during the assembly of the inner dark matter halo. Stellar discs retain their specific angular momentum since they are comprised of stars formed mainly after turnaround, from
gas that mostly preserves the high specific angular momentum it acquired by tidal torques during the linear growth of the halo. 
Since the specific angular momentum loss of the stars is tied to the galaxy's morphology today, it may be possible to use our results to predict, statistically, the assembly history of a halo given the morphology of the galaxy it hosts.
\end{abstract}


\section{Introduction}

In the standard model of structure formation, gravity drives the evolution of the dominant form of matter $-$dark matter$-$ to form self-bound structures (haloes).
The protohaloes gain angular momentum through tidal torques from their environment until maximum expansion (turnaround), and subsequently collapse into
virialized structures that preserve their angular momentum \citep{Doro1970,White1984,Catelan1996a,Catelan1996b}. In the early stages of galaxy formation, the dynamics of baryonic matter is driven by 
the gravity of the dark matter, closely following the dynamical evolution of the protohaloes. Therefore, it is expected that the protogalaxy will track the specific angular momentum
of the host halo until dissipative baryonic processes that transfer mass and/or angular momentum decouple the evolution of the galaxy from that of the host halo. 

The correspondence between the specific angular momentum of the galaxy and its host halo is complex because it depends on the detailed loss and gain of angular
momentum as the galaxy is assembled. Since the mass and specific angular momentum of observed galaxies are closely related to their morphology \citep{Fall1983}, understanding this
correspondence potentially enables us to infer the assembly history of a halo given the observed morphology of the galaxy it hosts. 

Disc galaxies are thought to be formed by the combination of two main modes of gas acquisition: (i) {\it isotropic accretion} from the condensation of a hot gas 
corona with a long cooling time that is shock-heated to the virial temperature of the halo throughout its assembly \citep{White1978,White1991}, and (ii) {\it anisotropic accretion} from cold gas with a short cooling time that flows directly onto the centre of the dark matter halo without being shock-heated \citep{White1991,Fardal2001,Keres2005,Dekel2009}. In the former, the galactic disc is formed as the corona loses energy while conserving angular momentum and
therefore preserving the affinity with the primordial angular momentum distribution of the host halo \citep{Fall1980}. In the latter, the disc is formed from
high angular momentum gas accreted through cosmic filaments surrounding the host halo, which contrary to the corona, never reaches hydrostatic equilibrium in the host halo, and is therefore linked less strongly to its primordial angular momentum distribution \citep[e.g.][]{Stewart2013}. 

On the other hand, spheroids (bulges and elliptical galaxies) are dispersion-supported systems with low angular momentum which are commonly 
thought to be formed through mergers of galaxies (mixing stars with distinct angular momenta from two or more galaxies, \citealt{Toomre1977}), internal secular processes in the disc (instabilities, for a review see \citealt{Kormendy2004}) and/or by continuous gas infall with misaligned angular momenta \citep{Sales2012}. In most of these processes, the majority of the angular momentum of the protogalactic fragments is transferred 
outside of the spheroid remnant. It is then expected that for those galaxies where the main channel of spheroid growth is galaxy mergers, their assembly is tightly related to the merger history of the parent host haloes \citep[a cornerstone assumption in semi-analytic and semi-empirical methods, e.g., ][]{Kauffmann1993,Bower2006,Croton2006,Hopkins2010,Zavala2012}.

Elucidating the intricate relation between the specific angular momentum evolution of the dark matter halo and the galaxy within is a challenge worth pursuing since it would ideally 
enable us to obtain a snapshot, at least in a statistical sense, of the assembly of the dark matter haloes from the observed kinematics of their galaxies. This task is particularly difficult 
given the complexities of the different baryonic processes at play. With the advent of a new generation of hydrodynamical simulations, we can attempt to address these challenges, since it is now possible to model the complexity of galaxy formation/evolution in a full cosmological setting  
(e.g. the recent state-of-the-art simulations, Horizon-AGN \citealt{Dubois2014}, Illustris \citealt{Vogelsberger2014} and EAGLE \citealt{Schaye2015}). This encourages the exploration of the link between halo and galaxy assembly based on a physical modelling which, although it carries a large number of free parameters, is able to obtain reasonable galaxy morphologies. It is worth noting that the free parameters in these simulations are mainly calibrated by matching the present day stellar mass function, which does not automatically guarantee a good match to the morphologies. This requires the consideration of other observables related to morphology when calibrating the parameters; in EAGLE for example, this was done by looking at $z=0$ galaxy sizes \citep{Crain2015}.

A substantial number of studies have examined the halo-galaxy angular momentum link using numerical simulations. The following are some of the findings relevant to this article. Pioneering works, although subject to numerical artifacts due to limited resolution, already showed that suppressing the transfer of angular momentum from the baryons to the outer dark matter halo is a prerequisite for the formation of realistic disc galaxies \citep{Weil1998,Navarro2000,Thacker2001}. The key is efficient stellar feedback at early times, which reheats the gas into an extended hot reservoir, avoiding the strong merging activity in the centre of the halo \citep[e.g.][]{Sales2010,Brook2011}. 
A disc galaxy can then form from the gas that subsequently cools, conserving its angular momentum. In this scenario, the evolution of the specific angular momentum of the galaxy follows closely that of the entire halo, confirming
the classical picture of disc formation \citep[][]{Fall1980,Mo1998}, as was explicitly shown by \citet{Zavala2008}. Angular momentum can however, be redistributed or lost from the disc altogether in subsequent mergers and/or disc instabilities, while a new disc can be generated from further gas accretion \citep[e.g.][]{Governato2009}. The morphology of a galaxy is thus a transient state of galaxy evolution. Other studies  \citep[e.g.][]{Danovich2015} have shown that the {\it anisotropic mode} of accretion inherently carries more angular momentum than the {\it isotropic mode} (and therefore the accreting gas carries higher spin than the dark matter halo), making it an effective avenue of disc formation that complicates (complements) the classical picture.

One important aspect of the halo-galaxy angular momentum connection is the spatial region of the halo that is linked most strongly with the galaxy. Numerical simulations have shown that the angular momentum vectors of galactic discs are better aligned (albeit moderately) with that of the inner regions of their haloes (within $\sim10\%$ of their virial radius), than with the angular momentum vectors of entire haloes \citep[within their virial radius, e.g.][]{Bailin2005,Bett2010,Vell2015}. This alignment is the result of a combination of a common origin and subsequent evolution of the protogalaxy and the inner protohalo, and/or the dynamical response of the inner halo to the assembly of the disc. 

In this work we revisit the evolution of the specific angular momentum of galaxies and their dark matter haloes following the approach of \citet{Zavala2008}, with the relevant advantage of using a large sample of simulated galaxies from the EAGLE project, whose global properties (e.g. the stellar mass function and the distribution of galaxy sizes) are in good agreement with basic observables of $z=0$ galaxies. Our goal is to establish whether or not, despite the complexities of baryonic physics, there is a prevalent connection between the provenance and assembly of the dark halo and the destiny of the galaxy it hosts, in terms of specific angular momentum evolution and present morphology. 

This paper is organised as follows. In Section \ref{sec_sim} we provide a brief description of the simulation we use, 
define the different components of the galactic/halo system that we analyse, and
establish a simple dynamical method of morphological classification. In Section \ref{sec_AM_evolution}, we analyse the specific angular momentum evolution of the different components of the galactic/halo system, while in Section \ref{sec_AM_morphology} we discuss the connection between specific angular momentum loss and morphology (among the different components). Finally, in Section \ref{sec_Conclusions} we present a discussion of our results and give our conclusions.

\section{The EAGLE simulation}\label{sec_sim}

We use one of the main simulations of the Virgo Consortium's EAGLE project \citep{Schaye2015,Crain2015}
to perform our analysis. This simulation is labeled Ref-L100N1504 and was performed
in the context of a Planck cosmology \citep{Planck2014} with parameters:
$\Omega_m=0.307$, $\Omega_b=0.04825$, $\Omega_{\Lambda}=0.693$, $h=0.6777$,
$\sigma_8=0.8288$ and $n_s=0.9611$; where $\Omega_m$, $\Omega_b$ and $\Omega_{\Lambda}$ are the
contribution from matter, baryons and the cosmological constant to the mass/energy density
of the Universe, respectively, $h$ is the dimensionless Hubble
parameter at redshift zero, $\sigma_8$ is the rms amplitude of linear mass fluctuations
in $8h^{-1}~$Mpc spheres at redshift zero, and $n_s$ is the spectral index of the primordial
power spectrum. The simulation has a box size of 100 comoving Mpc on a side and follows the evolution
of $1504^3$ dark matter particles and an initially equal number of baryonic particles. The dark matter particle mass is $9.7\times10^6$M$_\odot$, while the initial gas particle mass is
$1.81\times10^6$M$_\odot$. The gravitational softening length (Plummer-equivalent) was fixed in comoving units for $z\ge2.8$  ($\epsilon=2.66$~kpc) and in proper units thereafter
($\epsilon=0.70$~kpc). 

\begin{figure*}
\begin{tabular}{|@{}l@{}|}
\includegraphics[width=6.5cm,height=6.5cm]{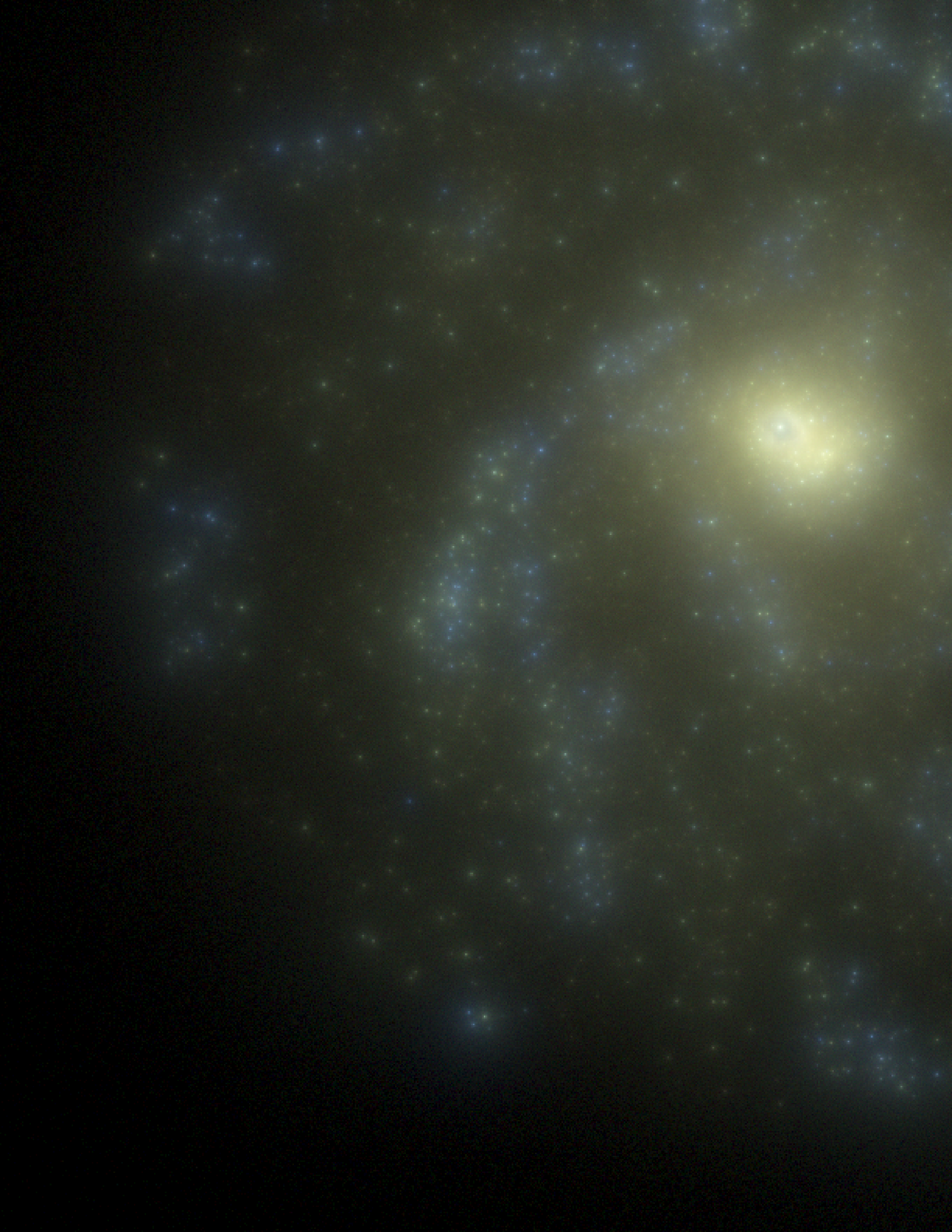} 
\includegraphics[height=6.5cm, width=8.25cm, trim=0cm 0.6cm 0cm 1.2cm, clip=true]{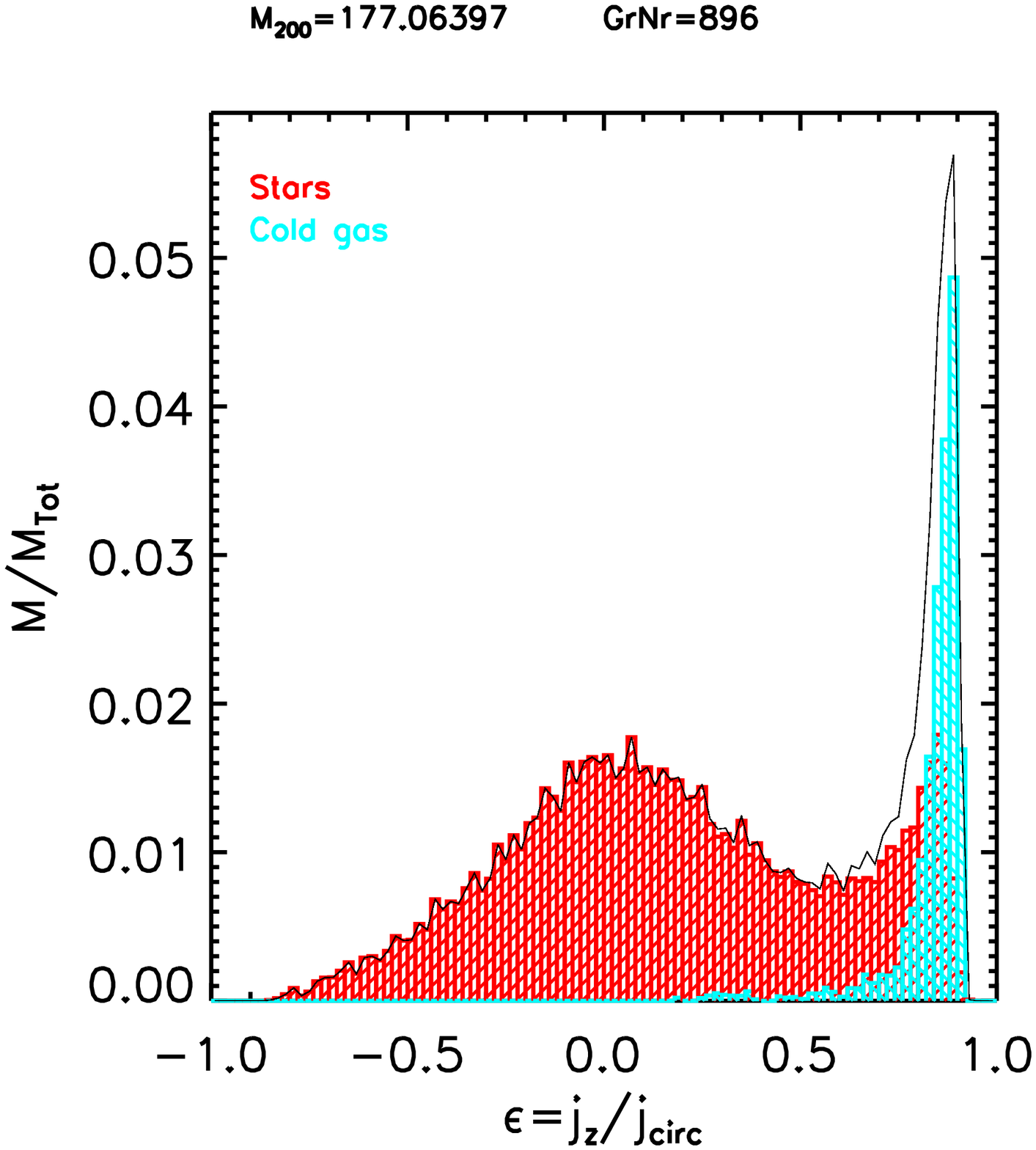}
\end{tabular}
\caption{Example of one of the galaxies analysed from EAGLE Ref-L100N1504. {\it Left panel}: The image (60 kpc on a side) was created with the radiative transfer code
SKIRT \citep[][Trayford et al. in prep.]{Camps2015}. It shows the stellar light based on monochromatic u, g and r band SDSS filter means, and accounting for dust extinction.  
Only a subset of the particles shown in the image were analysed for this paper (see Section \ref{sec_lagrangian}). {\it Right panel}: orbital circularities of the cold baryons (stars in red, gas in cyan, stars + gas in black) for the galaxy shown on the left. A simple bulge disk decomposition based on kinematics can be performed
by defining the bulge mass as twice the mass to the left of the circularity peak (i.e., the bulge is a left-to-right reflection around $\sim$zero circularity).} 
\label{example_image}
\end{figure*}

EAGLE was run using a modified version of the {\tt GADGET 3} code (last described in \citealt{Springel2005}). 
It uses a set of numerical methods, referred to as {\tt ANARCHY} (Dalla Vecchia, in preparation; see also \citealt{Schaller2015}, and Appendix A of \citealt{Schaye2015}) with its core hydrodynamics 
based on the pressure-entropy formulation of smoothed particle hydrodynamics (SPH, \citealt{Hopkins2013}), and the time step limiters of \citet{Durier2012}. The baryonic physics implementation includes element-by-element cooling, and
heating from the photo-ionisation by a spatially uniform time-dependent UV/X-ray background \citep{Wiersma2009a}, stochastic star formation with a star formation rate that depends explicitly on pressure (\citealt{Schaye2008}; with a star formation density threshold that is metallicity dependent; \citealt{Schaye2004})\footnote{See the last paragraphs of Section \ref{sec_morpho} for a discussion on the treatment in EAGLE of the unresolved multiphase interstellar medium, and the impact of this treatment on this work.}, a Chabrier IMF, time-resolved stellar mass loss \citep{Wiersma2009b}, thermal energy feedback associated with star formation \citep{DV2012}, and the growth of black holes. The latter grow from initial seeds of $10^5h^{-1}$~M$_\odot$ via mergers and Bondi gas accretion, with a modification for angular momentum of the surrounding gas \citep{Rosas2015}, and AGN feedback (thermal and stochastic). See \citet{Schaye2015} for a complete description of the physical modeling in EAGLE, and \citet{Crain2015} for information on the calibration of the reference simulation we use here (EAGLE Ref-L100N1504).

Haloes and galaxies are identified in EAGLE using the {\tt SUBFIND} algorithm \citep{Springel2001,Dolag2009}, firstly identifying haloes with the Friends-of-Friends (FoF) algorithm (with a linking length of 0.2 times the mean interparticle separation), then creating a hierarchy of gravitationally self-bound substructures for each FoF halo. The most massive subhalo of this
hierarchy is the main or distinct halo that hosts a central galaxy, while some of the most massive subhaloes host satellite galaxies.

\subsection{The Lagrangian components}\label{sec_lagrangian}

We aim to track the temporal evolution of the population of central galaxies identified at $z=0$. For this purpose we examine the 3928 most massive main haloes (and their central galaxies) in the EAGLE Ref-L100N1504 simulation. These correspond to haloes with a virial mass\footnote{Defined as the enclosed mass at the radius within which the mean density is 200 times the critical density of the Universe.} $M_{200}\ge2\times10^{11}$M$_\odot$, and central galaxies with a stellar mass $M_{\ast}\ge4\times10^8$M$_\odot$. For our analysis we define three Lagrangian components that determine the galactic system at $z=0$ (following \citealt{Zavala2008}):
\begin{itemize}
\item {\it The whole dark matter halo}: Comprised of all dark matter particles within the virial radius of the FoF halo. 
\item {\it The inner dark matter halo}: Comprised of all dark matter particles within $10\%$ of the virial radius of the FoF halo.
\item {\it The galaxy}: Comprised of cold baryons within the inner dark matter halo: all star particles and gas particles with $\rho_g > 7\times10^{-27}$g cm$^{-3}$ (an overdensity threshold with respect to the cosmic baryon density of 
$\sim1.7\times10^4$)\footnote{This is a density threshold that was chosen to select the radiatively cooled ($T\sim10^4$~K) dense gas. Since we are only considering the gas within $0.1r_{200}$, most of the gas is in the condensed and cool regime. The specific value of this threshold does not impact our results in any significant way.}.
\end{itemize}

At all epochs, this set of particles represents the Lagrangian region that will eventually become the galaxy (halo) at $z=0$. Notice that due to the particle-based
nature of the SPH implementation in EAGLE, a star particle at $z=0$ represents a fully evolved stellar population that, as it is tracked back in time, is transformed into 
a progenitor gas particle. This process can be followed unambiguously since each star particle carries a unique identification number, which is inherited from the gas particle from which it formed. Due to stellar mass losses, the masses of the Lagrangian star/gas particles are not conserved during the evolution.
 {\it Unless stated otherwise, when we refer to any of the galactic components in any given epoch, we are always referring to the Lagrangian set of particles identified at $z=0$.}

\subsection{Morphological classification based on kinematics}\label{sec_morpho}

We use a ``morphological'' classification of the galaxies at $z=0$ based on the kinematics of the galaxy (today's cold baryons). Specifically, we use the distribution of circularities $\epsilon$, of the baryons in the galaxy as a proxy for morphology:
\begin{equation}
\epsilon=\frac{j_z}{j_{\rm circ}(E)},
\end{equation} 
where $j_z$ is the specific angular momentum of each particle projected in the direction of the total angular momentum of the cold baryonic component, and $j_{\rm circ}(E)$ is the 
specific angular momentum corresponding to a circular orbit with the same binding energy $E$ of the particle. 
The positions and velocities of the particles are measured with respect to the center of mass frame of the cold baryons. 

The presence of a rotationally supported disc and/or a dispersion-supported bulge in a given galaxy is easily identified by examination of the distribution of circularities. The stars comprising
the bulge are distributed about $\epsilon\sim0$, while those of the disc exhibit much higher values ($\epsilon\sim1$). An example of the spatial distribution and circularities of one of the EAGLE galaxies we analyse is shown in Fig.~\ref{example_image}. The presence of both stellar components is evident in this galaxy; a cold gaseous disc is clearly visible as well. Thus, we define a simple kinematic measure of morphology by defining the mass of the spheroid component to be $M_{\rm bulge}=2M(\epsilon<\epsilon_{\rm peak})$, where 
$\epsilon_{\rm peak}$ is the peak near $\epsilon=0$ \citep{Abadi2003}\footnote{In the cases where it is difficult to separate the peaks corresponding to the bulge and the disc, we explored the
suggestion of \citet{Martig2012} to disentangle the peaks using the circularity of only the central stars, within $\sim2\%$ of the virial radius. The latter corresponds to just a few times the softening length for the lowest mass haloes we analysed, which are thus affected by resolution at these small scales. We therefore extend this central radius and explored the inner regions between $2-5\%$ of the virial radius, but at the end, we found that 
this step does not make a significant difference in our results compared to the case where we simply define the spheroid using $\epsilon_{\rm peak}\sim 0$. We therefore keep the latter simpler definition.}.

\begin{figure*}
\begin{tabular}{|@{}l@{}|}
\includegraphics[height=8.0cm,width=8.75cm]{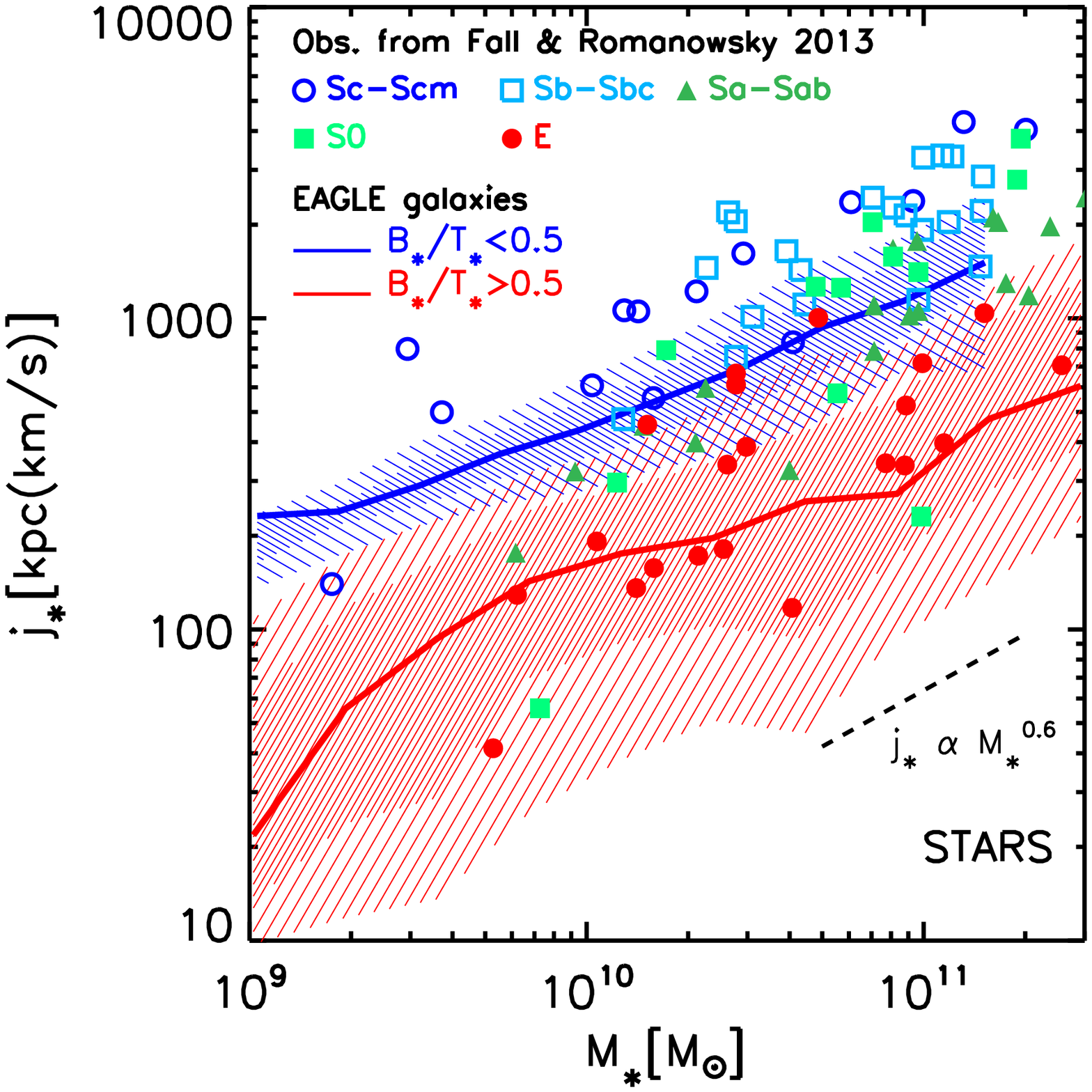}
\includegraphics[height=8.0cm,width=8.75cm]{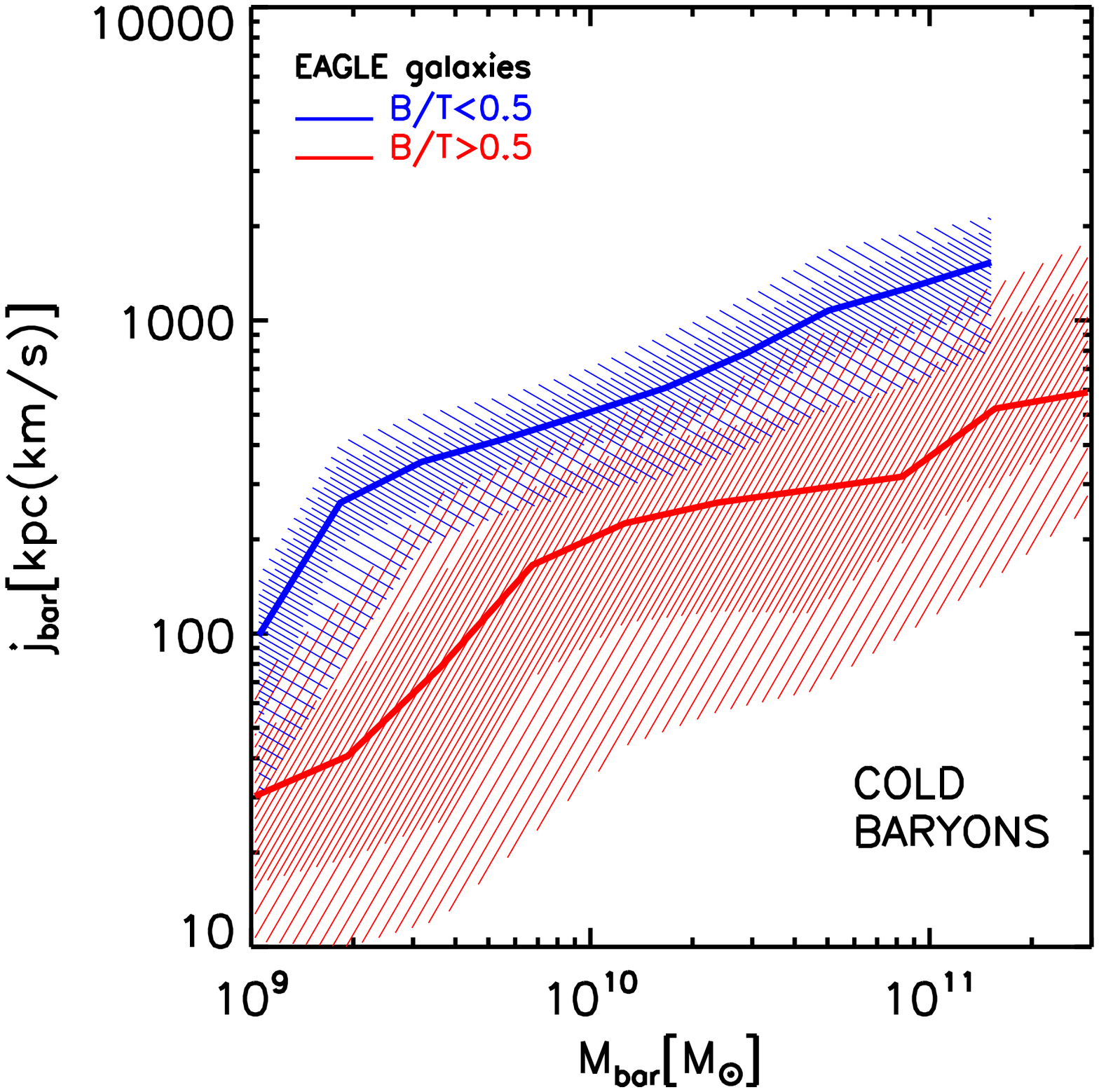} 
\end{tabular}
\caption{Specific angular momentum as a function of mass at $z=0$ (stars on the left, cold baryons on the right). The sample of simulated galaxies has been divided in two types according to the value of the bulge-to-total mass ratio (defined from the circularity distributions of the stars in the left and from all cold baryons in the right): red (bulge-dominated, $B/T\geq0.5$) and blue (disc-dominated, $B/T<0.5$). The thick lines show the medians of the distributions of each type whereas the hashed areas encompass the $10\%-25\%$, $25\%-75\%$ and $75\%-90\%$ regions. The symbols on the left panel show the observations presented in \citet{Fall2013} separated in different morphological types. The dashed line on the left shows the approximate scaling of the observed relation that the discs of spirals and ellipticals follow, $j_{\ast}\propto M_{\ast}^{0.6}$ \citep{Fall2013}. }
\label{AM_Fall} 
\end{figure*}

Having defined the total (stars + cold gas) mass in the spheroid (bulge), a total bulge-to-total mass ratio ($B/T$) can be assigned to each galaxy. We repeat the previous procedure but using {\it only} the stars in the galaxy, starting from the definition of circularity, i.e., using the specific angular momentum vector of the stars, as opposed to that of the baryons, as the axis of rotation. This defines a stellar bulge-to-total mass ratio ($B_\ast/T_\ast$). We found that using a simple cut in the bulge-to-total stellar mass ratio divides the galaxy population into bulge-dominated ($B_\ast/T_\ast\geq0.5$) and disc-dominated galaxies ($B_\ast/T_\ast<0.5$), with a morphological mixture of $55\%/45\%$ for all galaxies with $M_\ast\geq10^{10}$M$_\odot$. This is a mixture that has perhaps too many bulge-dominated galaxies according to observations, which at these high stellar masses have a morphological mixture with a range between 
$\sim20\%/80\%$ (\citealt{Fisher2011} from a local volume-limited sample with $\sim$ 100 galaxies) and $\sim45\%/55\%$ (\citealt{Gadotti2009} from a volume-limited sample of $\sim1000$ local galaxies); see the top right panel of
\citet{Zavala2012} for a comparison of these observational samples. We remark however, that our kinematic morphological classification is not directly related to observational classifications, which are based on surface brightness profile fitting. Indeed, \citet{Scannapieco2010} found that kinematical classifications typically overestimate the $B/T$ ratio by large factors relative to the one found with photometric decompositions.

Despite this important caveat, the angular-momentum-mass-morphology connection is shown in the left (right) panel of Fig.~\ref{AM_Fall} were we plot the total specific angular momentum of the stars (cold baryons) as a function of stellar (cold baryons) mass for our sample of bulge- and disc-dominated galaxies (red and blue, respectively) in EAGLE at $z=0$. We show the median (thick line), and the 
$10\%-25\%$, $25\%-75\%$ and $75\%-90\%$ regions that divide each distribution (hashed areas).
The observational data presented in \citet{Fall2013} is shown in the left divided according to morphological type as shown in the legend (see \citealt{Romanowsky2012} for details on how the specific angular momentum is estimated from observations for the different galaxy types). The dashed line on the left panel shows the observationally inferred relation $j_{\ast}\propto M_{\ast}^{0.6}$ for the discs of spirals and ellipticals \citep{Fall2013}. This scaling agrees well with the theoretical analysis by \citet{Catelan1996a}. 
The simulated galaxies follow this scaling and seem to be within the observed distributions as a whole, although there is a vertical offset indicating a larger specific angular momentum for the observed galaxies at a fixed stellar mass. Our kinematic morphological classification (albeit simply divided in two broad galaxy types) seems to separate the galaxies in a way that is consistent with observations, i.e. disc-(bulge-)dominated galaxies corresponding roughly to the region of morphological types Sc$-$Sa (E). We note that \citet{Genel2015} have also recently reported a good agreement between observations and their simulated galaxies from the Illustris project in this angular-momentum-mass-morphology relation. 

The right panel of Fig.~\ref{AM_Fall} shows the baryonic relation (stars plus gas), which shows similar correlations although there is a clear drop of specific angular momentum at lower masses ($M_{\rm bar}\lesssim5\times10^{9}$M$_\odot$, also visible but less clearly for bulge-dominated galaxies in the left panel). This is likely related to the more prominent role of gaseous discs for low-mass galaxies, which might suffer from limited resolution. 
Numerical artefacts could be important at low masses\footnote{For instance, \citet{Furlong2015} found that the fraction of galaxies in EAGLE with specific star formation rate below 0.1~Gyr$^{-1}$ (passive) artificially rises below $M_\ast<3\times10^9$M$_\odot$. This is a consequence of inadequate sampling of the star formation rate, which can indirectly impact morphology as well since without a good sampling, is not possible to form stars in a well-defined disc.}, although galaxies with $M_{\ast}\sim10^{10}$M$_\odot$ have $\sim5\times10^3$ star particles, which should be enough to sample the circularity distributions. More important is the fact that in EAGLE, the cold star forming gas phase is not resolved or followed. For the warm interstellar medium (ISM), $T\sim10^4$K and $n_H\sim0.1$cm$^{-3}$, thus, the Jeans length is $\lambda_{J}(\rm ISM)\sim1$~kpc, whereas for the cold ISM, the temperatures (densities) are typically 100 times lower (higher) implying a Jeans length of the order of $40$~pc (larger if the pressure is dominated by turbulence). With a softening of 700 pc, the simulation we use is only able to marginally resolve the warm ISM but it cannot resolve the cold ISM. 

To circumvent this, the EAGLE project uses a star formation threshold, which is imposed exactly at the density above which a cold ISM phase is expected to form,
and a temperature floor given by an effective equation of state $T\propto\rho^{1/3}$.
Stars are formed from gas with a density above this metallicity-dependent density threshold and with temperatures near this temperature floor. The latter introduces a thermal pressure with a corresponding Jeans length that varies as: $\lambda_J\propto (T/\rho)^{1/2}\propto\rho^{-1/3}$. The weak scaling with the density implies that even for high-density gas in star forming discs in the simulation, the Jeans length associated with the temperature floor is not far from that of the warm ISM $\lambda_J(\rm ISM)\sim1$~kpc.
Gaseous discs with a characteristic scale $\lambda\lesssim\lambda_J$ will be strongly affected (they become broader) by this ``subgrid'' thermal pressure, which is required due to the final resolution of the simulation. Examination of Fig.~7 of \citet{Crain2015} indicates that 
most of the stars in the simulation are formed from gas with densities above the peak of the distribution shown in that figure ($n_H\sim0.3$~cm$^{-3}$) and below $n_H=100\times n_H(\rm ISM)=10~$cm$^{-3}$. Only $\sim25\%$ of the stars are born with even higher densities. Therefore, we expect that the bulk of the stars in the simulation will have a ``subgrid'' pressure with associated Jeans length somewhere in the range $0.4$~kpc~$\lesssim\lambda_J\lesssim 1.4$~kpc\footnote{We note that the effective pressure in the real ISM will be of similar order, since real disks are also in near vertical hydrostatic equilibrium. So the subgrid pressure in EAGLE takes into account unresolved sources of ``pressure'' such as turbulence.}. Since the median half-mass radius of $M_{\ast}\sim10^{10}$M$_\odot$ galaxies is $\sim3$~kpc \citep[see Fig. 3 of][]{Crain2015}, galaxies with lower masses are expected to have their stellar orbits affected by an artificial velocity dispersion, which potentially impacts the circularity distributions (i.e., the $B/T$ ratios), and also the total specific angular momentum. 
Instead of making a full analysis of the impact of this physical resolution limit in our work, we have tested whether our main results (statistical in nature) change significantly in the galaxies with the lowest stellar masses. We have found that galaxies with $M_\ast\le7.5\times10^9$M$_\odot$ show some departures in the correlations we 
present in this work (most significantly in Fig.~\ref{bt_vs_am_loss_baryons}), which is roughly the stellar mass threshold where we anticipated resolution issues, given the discussion above. Therefore, henceforth we only consider
galaxies above this threshold, which reduces our sample to 2488 galaxies.

\begin{figure*}
\begin{tabular}{|@{}l@{}|}
\includegraphics[height=8.5cm,width=8.5cm]{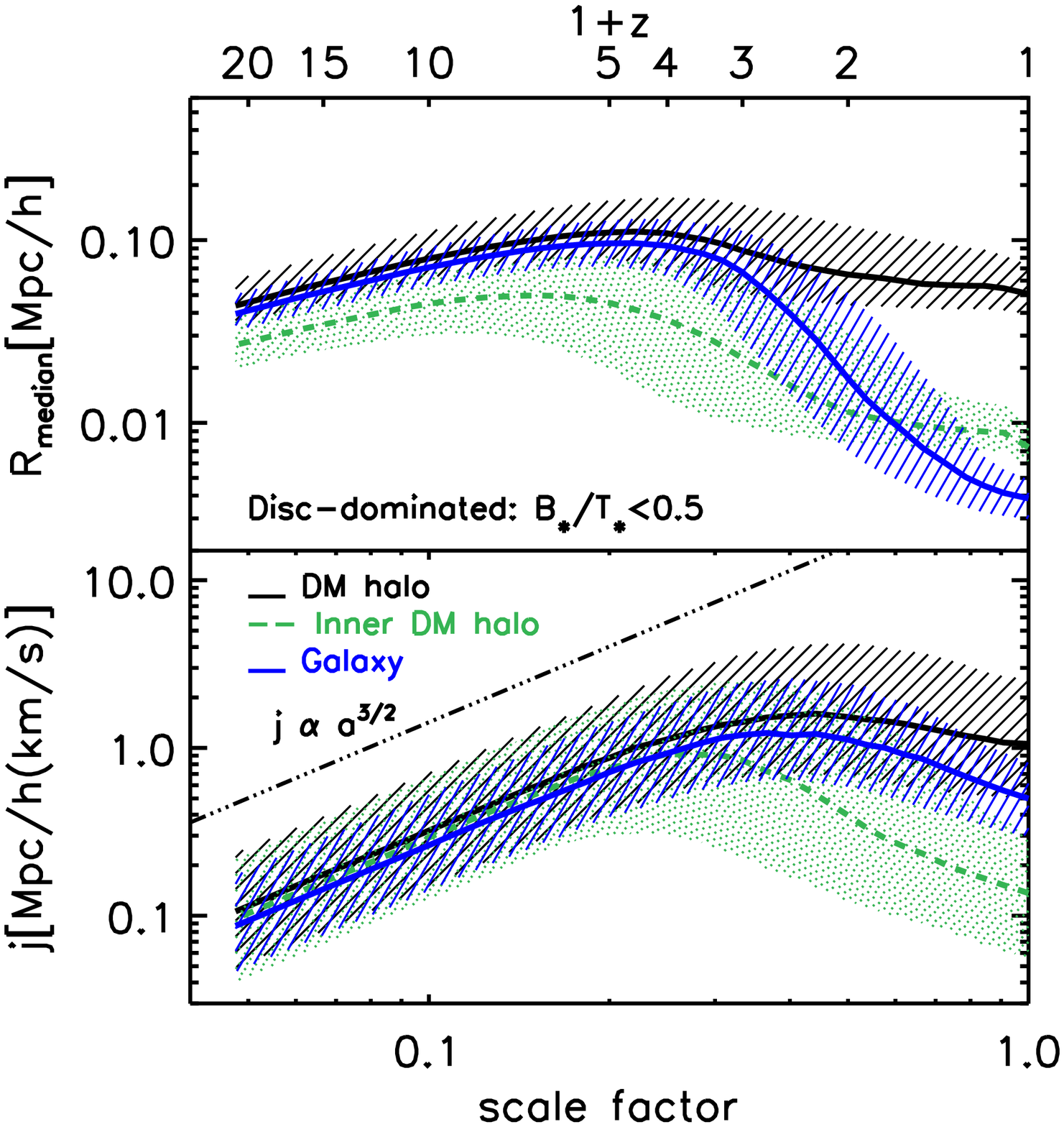}
\includegraphics[height=8.5cm,width=8.5cm]{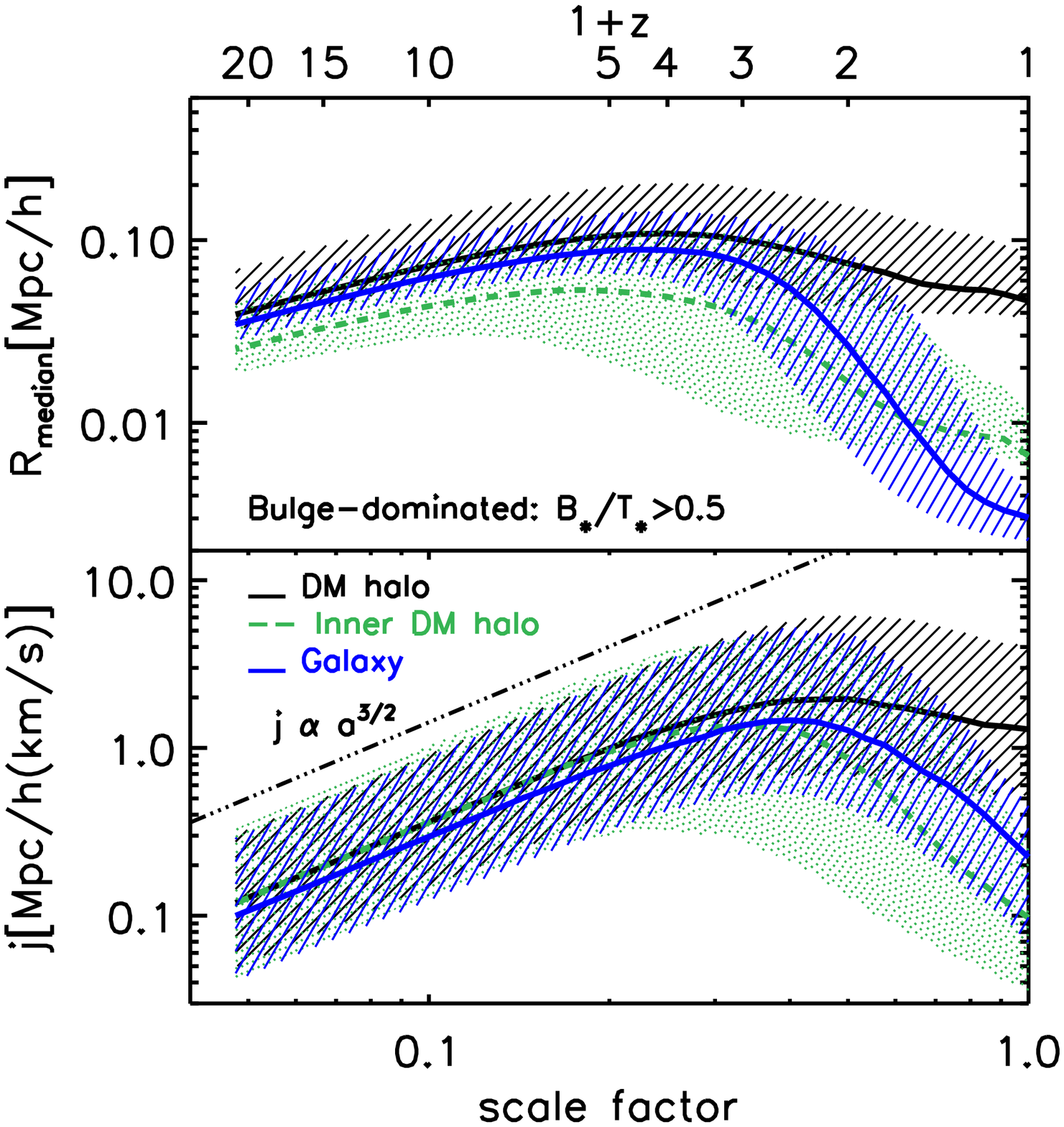} 
\end{tabular}
\caption{Evolution of the median radius (top) and the specific angular momentum (bottom) of the Lagrangian set of particles that define the three galactic components at $z=0$ (see Section \ref{sec_lagrangian}):
dark matter halo (black solid), inner dark matter halo (green dashed) and the galaxy (cold gas + stars, blue). The thick lines show the median of the distributions whereas the hashed areas
cover the  $\pm1\sigma$ regions. The galaxy sample is divided into disc-dominated (left) and bulge-dominated (right) galaxies according to their $z=0$ stellar bulge-to-total mass ratios as shown in the legends. The dot-dashed lines show the prediction of specific angular momentum growth according to the linear tidal torque theory before turnaround \citep{White1984,Catelan1996a}.}
\label{AM_evolution} 
\end{figure*}

\section{Specific angular momentum evolution of the Lagrangian components}\label{sec_AM_evolution}

\begin{figure}
\centering
\includegraphics[height=8.25cm,width=8.25cm]{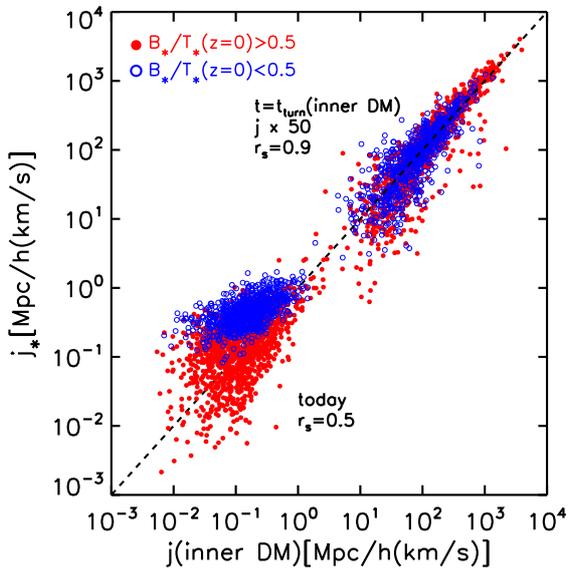} 
\caption{Correlation between the specific angular momentum of the stars and that of the inner dark matter halo (Lagrangian components) at two different times: today ($z=0$) and at the time of turnaround of the inner dark matter halo. For the latter epoch, the specific angular momentum is multiplied by 50 to avoid overlap between both epochs in the plot. The sample of  galaxies is divided into two galaxy types according to present-day morphology: bulge-dominated galaxies ($B_\ast/T_\ast\geq0.5$; solid red), and disc-dominated galaxies ($B_\ast/T_\ast<0.5$; open blue). The Spearman's rank correlation coefficient, $r_s$, is shown for each epoch.} \label{am_st_vs_dm10}
\end{figure}

The particles that define the three different (Lagrangian) galactic components at $z=0$ are tracked into earlier epochs to follow their specific angular momentum evolution. For each snapshot in the simulation, we compute the total specific angular momentum and, as a measure of size, the total median radius (relative to the time-dependent center of mass of the Lagrangian particles) for all particles in each component (both in physical units). We remark that in the case of the star and gas particles, the Lagrangian particles at any given epoch are {\it only} those which can be traced to $z=0$. Thus, when the specific angular momentum evolution is computed, a smaller (larger) number of star (gas) particles are used at higher redshifts compared to $z=0$, since star particles are continuously being born from gas particles.

At each epoch, we therefore have a distribution of sizes and total specific angular momentum for all the galaxies in our sample; from these distributions, we compute the median and $\pm1\sigma$ regions. This is shown in Fig.~\ref{AM_evolution} with the sample of galaxies divided according to their stellar 
bulge-to-total mass ratio at $z=0$:  bulge-dominated ($B_\ast/T_\ast\geq0.5$) galaxies to the right and disc-dominated ($B_\ast/T_\ast<0.5$) galaxies to the left. We note that this plot looks qualitatively similar if, instead of splitting according to $B_\ast/T_\ast$, we split the galaxy population according to stellar mass at $z=0$ with a dividing value of $3\times10^{10}h^{-1}$M$_\odot$. This is because this stellar mass roughly marks the transition between the dominion of disc-dominated (at lower masses) and bulge-dominated (at higher masses) central galaxies observationally \citep[e.g., see Fig. 3 of][]{Hopkins2009}.

The evolution of the size for the dark matter halo (as given by the median of the physical distance of all its particles relative to the center of mass) in Fig.~\ref{AM_evolution} follows closely the expectation from the spherical collapse model, with the system expanding, reaching a maximum and then collapsing to form the virialized halo today with a physical size roughly half of that at maximum expansion. Notice that the inner halo, having a higher overdensity than the whole halo, collapses earlier and shrinks to a size a factor of $\sim4-5$ smaller than at maximum expansion.

The evolution of the specific angular momentum of the whole dark matter halo follows the behaviour predicted by tidal torque theory until turnaround \citep[][dot-dashed line]{White1984,Catelan1996a}. Afterwards, the specific angular momentum is approximately conserved. The relatively small loss of specific angular momentum is related to the {\it arbitrary} boundary set by the virial radius. During the assembly of the halo, redistribution of angular momentum can actually place high angular momentum material outside the virial radius \citep{Donghia2007}. If the Lagrangian region of the whole halo is extended to $\sim3r_{200}$, then the specific angular momentum losses after turnaround are minimal. On the other hand, the inner dark matter halo loses $\gtrsim90\%$ of its specific angular momentum after turnaround. This significant loss is
a consequence of the transfer of angular momentum that takes place between the dark matter clumps that form the inner halo and the outer halo (first noted by \citealt{Frenk1985}, see also \citealt{Zavala2008}).

Today's cold baryons follow closely the dark matter halo until turnaround but then lose a significant fraction of their specific angular momentum, although not as much as the inner dark matter halo. Bulge-dominated galaxies ($B_\ast/T_\ast>0.5$)
have lost most of their specific angular momentum since turnaround ($\sim80\%$ for the median) and track the inner dark matter halo more closely (right panel of Fig.~\ref{AM_evolution}). Disc-dominated galaxies on the other hand, lose $50\%$ of their specific angular momentum (median) and track the behavior of the whole dark matter halo more closely.
Fig.~\ref{AM_evolution} seems to confirm the general expectation that disc-dominated galaxies today (predominant among low mass isolated central galaxies) were assembled with a lower relative loss (since turnaround) of specific angular momentum than bulge-dominated galaxies (predominant among massive isolated galaxies).

We remark that the specific angular momentum values in Fig.~\ref{AM_evolution} are unnormalised. This plot then shows clearly that before turnaround, the different components occupy very similar
Lagrangian regions (the inner halo and the galaxy being more concentrated), and are being torqued by the environment in a very similar way (this is clear in a statistical sense, but see Section \ref{sec_provenance}). In the following we concentrate on the evolution of the galactic components after turnaround.

In Fig.~\ref{am_st_vs_dm10} we compare the specific angular momentum of the stars and of the inner dark matter halo at two different epochs: at $z=0$ and at the epoch of turnaround for the inner dark matter halo. For the latter epoch, the specific angular momentum has been multiplied by 50 to show both epochs in the same plot without overlap.  Recall that although at two different epochs, the star (dark matter) particles are defined by the Lagrangian components identified at $z=0$. For dark matter, exactly the same particles appear at any given epoch, but for the stars, only a fraction of those identified at $z=0$ have formed at turnaround, the rest 
are still gas particles, and are thus absent from this plot. Therefore, at turnaround, Fig.~\ref{am_st_vs_dm10} shows the progenitor stars that are already in place at that epoch. 

The sample of 2488  central galaxies  has been reduced to 2005 to include only those galaxies that have at least 500 star
particles at turnaround so that the specific angular momentum measurements are less affected by poor particle sampling. Discreteness effects bias the measurement of the specific 
angular momentum upwards \citep[e.g.][]{Bett2007}. Thus, the threshold of 500 particles removes a tail of high-$j_{\ast}$ systems with low $j$(inner DM) (since these are typically the low-mass haloes affected by resolution, due to the natural mass-angular-momentum correlation)\footnote{We note that our threshold is slightly more conservative than that of \citet{Bett2007}, who suggest removing haloes with fewer than 300 particles.}. 
The galaxies are shown in Fig.~\ref{am_st_vs_dm10} according to their morphological classification at $z=0$ (using the stars only), which we described in Section \ref{sec_morpho}, disc- and bulge-dominated galaxies shown in open blue and solid red circles, respectively. 

As anticipated, 
there is a strong one-to-one correlation between the specific angular momentum of the stars and that of the inner dark matter halo at the epoch of turnaround (the Spearman's rank correlation coefficient is $r_s=0.9$), albeit with a considerable dispersion at low $j$ values. At that time, there is also no clear separation between the progenitor stars of the galaxies that will eventually become discs/spheroids. Deviations from this one-to-one relation become significant after turnaround. At $z=0$, the galaxies exhibit a more dispersed relation ($r_s=0.5$) that is, as expected, clearly separated in disc- and bulge-dominated galaxies, with the former having higher $j_\ast$ relative to the latter for a fixed $j$(inner DM). We focus now on understanding the main physical mechanisms governing the evolution of the specific angular momentum from the epoch of turnaround until today.

\subsection{The provenance of spheroids and discs}\label{sec_provenance}

\begin{figure*}
\begin{tabular}{|@{}l@{}|}
\includegraphics[height=8.0cm,width=8.0cm]{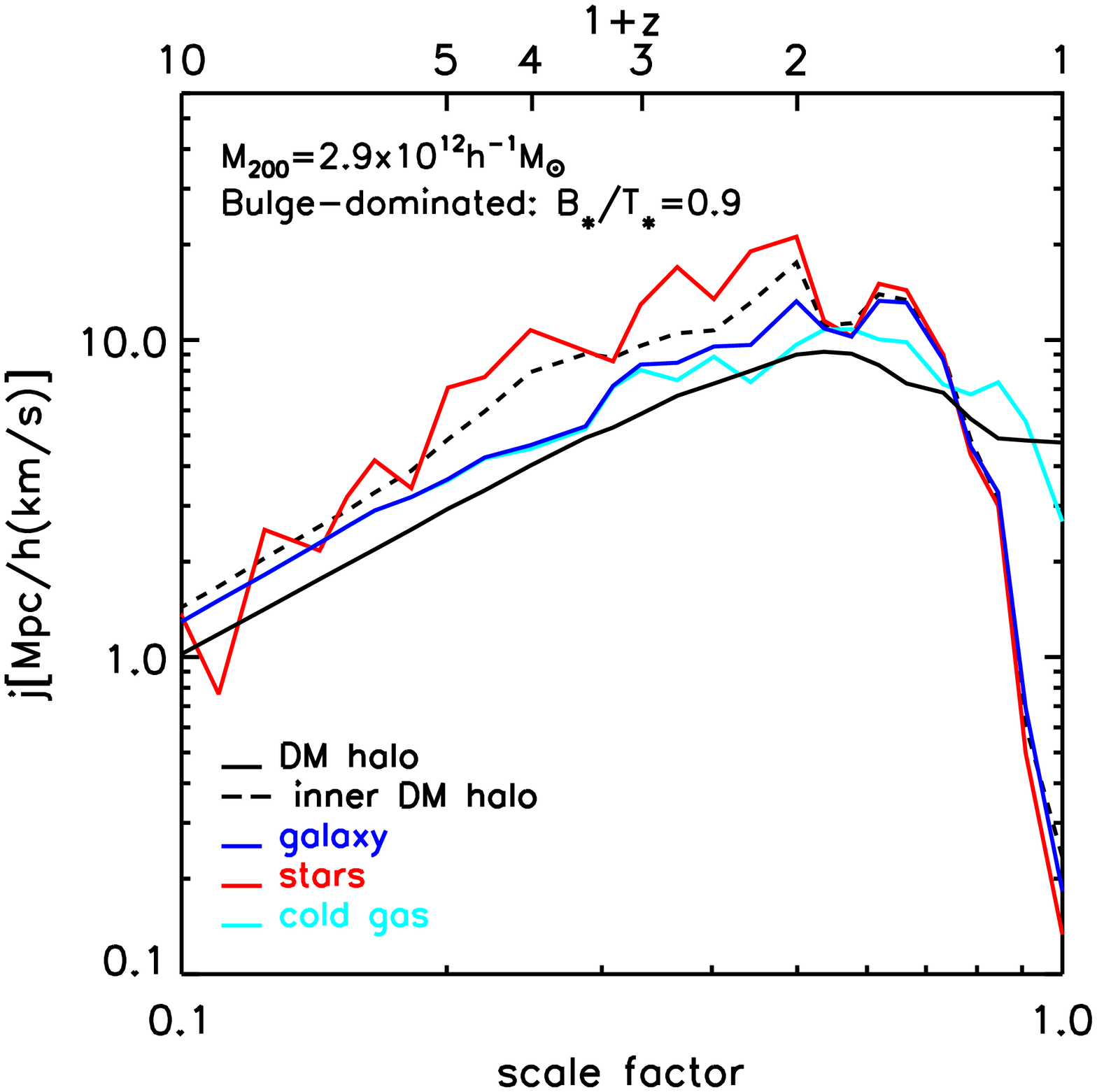}
\includegraphics[height=8.0cm,width=8.0cm]{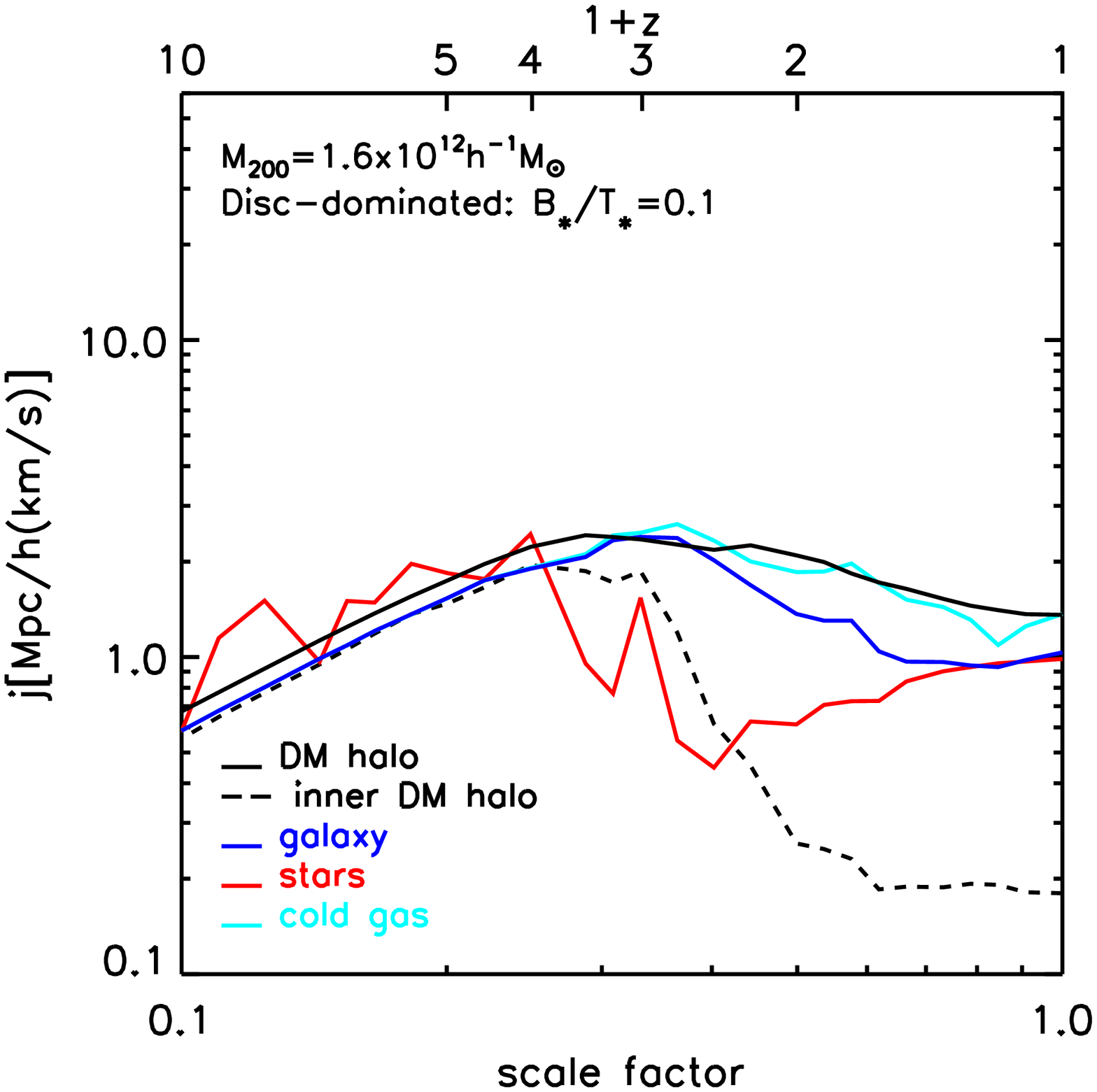} \\
\includegraphics[height=7.5cm,width=8.0cm, trim=1.4cm 0.4cm 0cm 1.0cm, clip=true]{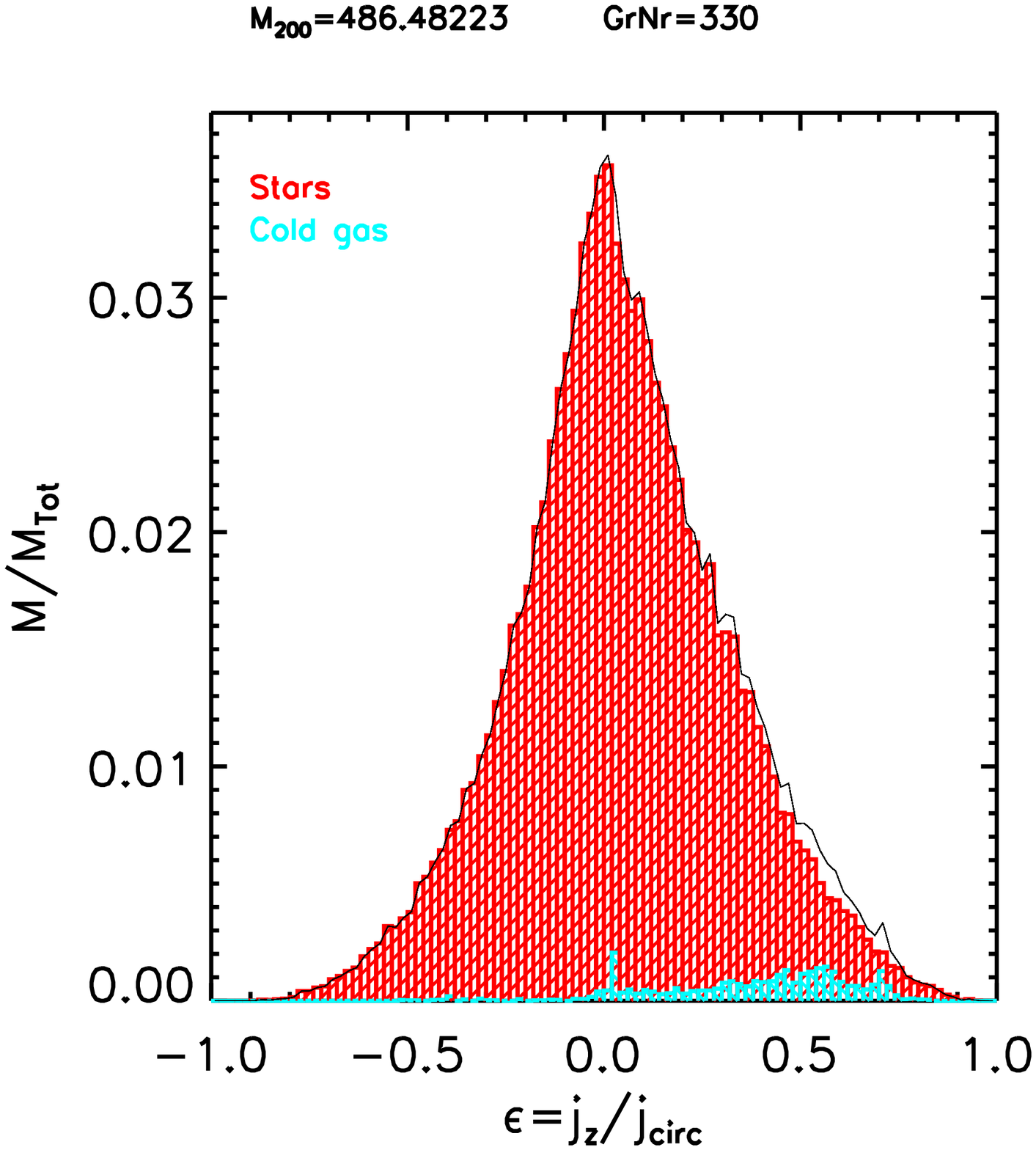}
\includegraphics[height=7.5cm,width=8.0cm, trim=1.4cm 0.4cm 0cm 1.0cm, clip=true]{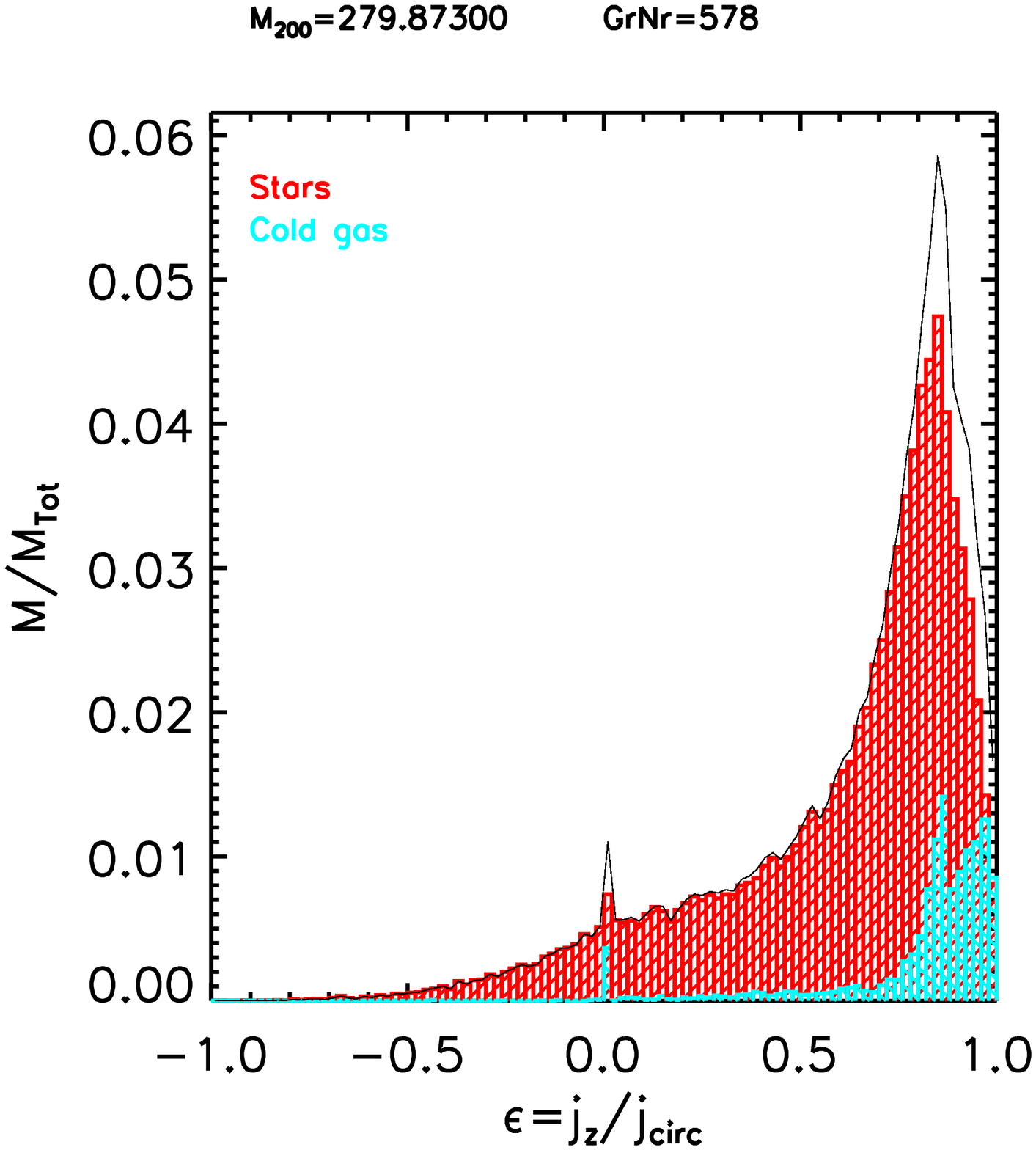} 
\end{tabular}
\caption{Examples of galaxies that today are bulge-dominated (left) and disc-dominated (right). {\it Top panels}: evolution of the specific angular momentum for the different Lagrangian components:
dark matter halo (solid black), inner dark matter halo (dashed black), stars (red), cold gas (cyan), and stars + cold gas (blue). Before the epoch when the inner dark matter halo reaches turnaround, most components track each other (typical behaviour for all galaxies). Afterwards, the correlation between the stars and the inner dark matter halo remains strong
for the bulge-dominated galaxy, while for the disc-dominated galaxy, the specific angular momentum of the stellar component increases after $z=1.5$ since young stars are being born from high angular momentum gas, which is uncorrelated with the inner dark matter halo. In many cases, like in these two examples, the gas follows the specific angular momentum of the whole dark matter halo relatively closely. {\it Bottom panel}: circularity distributions (at z=0) for the galaxies shown above (stars in red, cold gas in cyan).}
\label{AM_evolution_examples} 
\end{figure*}

Fig.~\ref{AM_evolution_examples} shows two representative galaxies that illustrate the connection between the evolution of the specific angular momentum (top panel) of the different Lagrangian components with the morphology of galaxies today (bottom panel). As we previously established (see Fig.~\ref{am_st_vs_dm10}), prior to the epoch of turnaround for the inner dark matter halo (at the peak of the dashed line in the top panel of Fig.~\ref{AM_evolution_examples}), all galactic components share a very similar specific angular momentum evolution; the evolutionary tracks are typically parallel to each other. There are however, some notable differences in the normalization of these tracks during the expansion phase. Before turnaround, the gas dominates the total specific angular momentum of today's cold baryons (compare cyan and blue lines in the top panel of Fig.~\ref{AM_evolution_examples}) and is typically higher in magnitude than the entire dark matter halo for most galaxies. The stars are subdominant during this time and have evolutionary $j_\ast$ tracks that are quite noisy, mostly because of poor particle sampling; the most massive galaxies are an exception since they can have a significant fraction of their stars already in place before turnaround. 


After turnaround the evolution changes drastically and is qualitatively different for galaxies that end up with different morphologies today. For the elliptical galaxy shown on the left of Fig.~\ref{AM_evolution_examples}, the bulk of the stars that define the galaxy today have already formed at turnaround, and are locked in the
merging subclumps of dark matter that will assemble the inner dark matter halo. Thus, they co-evolve sharing the same fate of transferring their orbital angular momentum to the outer halo. The similarity between the red (stars) and black (inner dark matter halo) dashed curves is remarkable. 
For the disc-dominated galaxy (right panels of Fig.~\ref{AM_evolution_examples}), the situation is quite different. Although the old stars and the inner dark matter halo track each other before and near turnaround, a dominant gaseous disc is already in place shortly after turnaround, 
which fosters the formation of a stellar disc of young stars, born with high specific angular momentum. 
It is this gaseous disc that is shaping the stellar morphology of this galaxy today (see bottom right panel of Fig.~\ref{AM_evolution_examples}).

Notice that for both galaxies, today's cold gas follows the specific angular momentum evolution of the whole dark matter halo relatively closely (for the elliptical galaxy there is a significant decline of specific angular momentum in the last stages since a large fraction of the gas is being funneled into the centre in a small burst of star formation). 


\subsection{Correspondence between the assembly of the central halo and today's cold baryons}\label{sec_correspond}

Although many galaxies fall within the categories of the examples shown in Fig.~\ref{AM_evolution_examples}, not all show correlations as tight and clean as illustrated there. To quantify how closely the baryonic components trace either of the dark matter components, we define the following measure of ``closeness":
\begin{equation}\label{goodness}
	Q^2 = \frac{1}{N(t_i>t_{\rm turn})} \sum_i\left[ {\rm ln}(j_1(t_i)) - {\rm ln}(j_2(t_i)) \right]^2, \ \ t_i>t_{\rm turn}
\end{equation}
where $j_1(t_i)$ and $j_2(t_i)$ are the specific angular momenta of the two components to be compared at a given time $t_i$ for the different snapshot outputs of the simulation. We only consider epochs {\it after} turnaround ($t_{\rm turn}$) of the inner dark matter halo of a given galaxy. Thus, $N(t_i>t_{\rm turn})$ is the number of snapshots we have for a given galaxy after turnaround. Defined in this way, $Q$ is similar to a goodness of fit test. If the two components track each other $Q\sim0$.

\begin{figure}
\centering
\includegraphics[height=8.0cm,width=8.0cm]{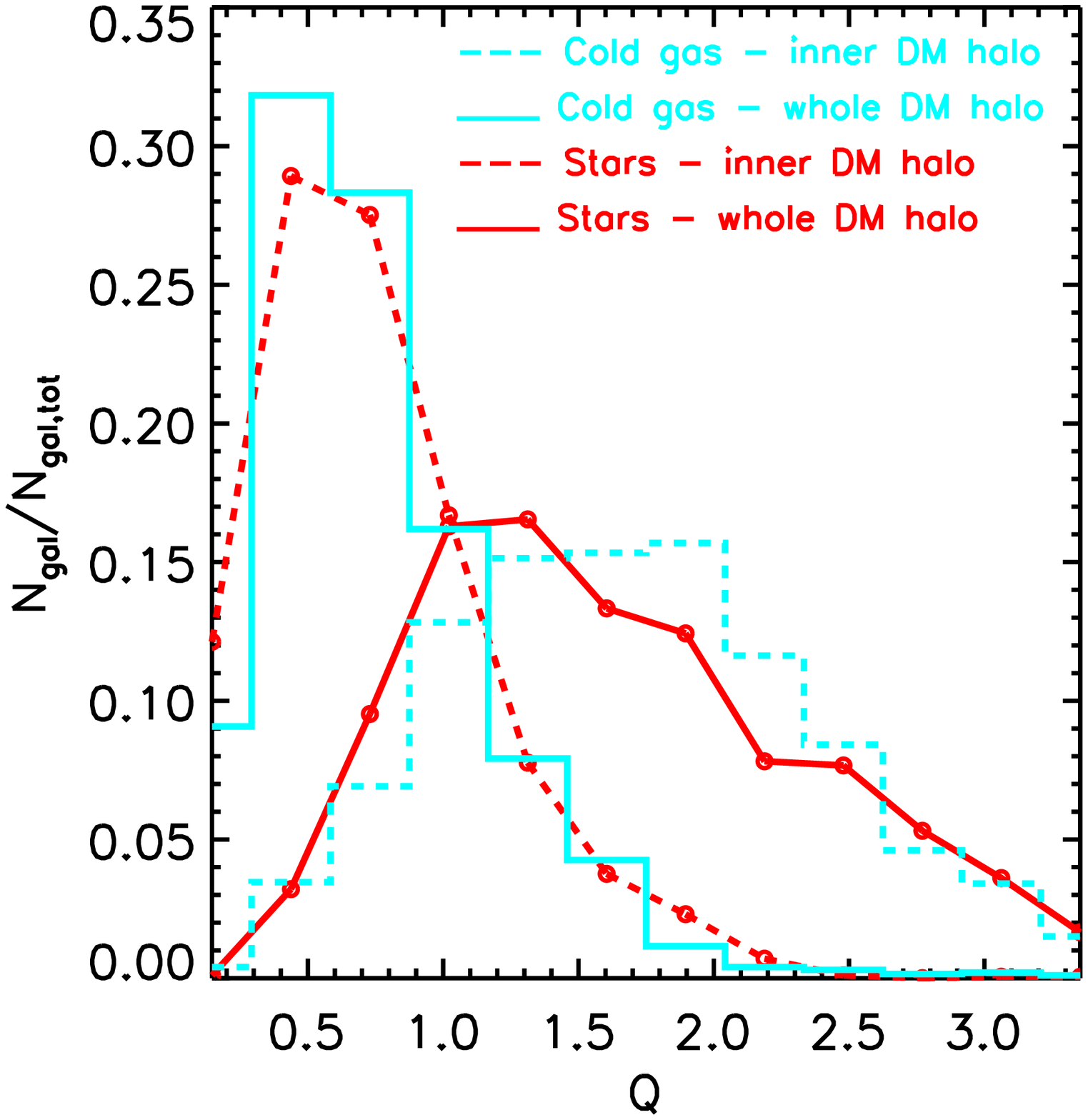} 
\caption{Normalised distribution of the ``closeness" ($Q$, Eq.~\ref{goodness}) between the specific angular momentum evolution of the different Lagrangian components measured after turnaround (of the inner dark matter halo): 
cold gas$-$dark matter halo (solid cyan), stars$-$dark matter halo (solid red), cold gas$-$inner dark matter halo (dashed cyan), and stars$-$inner dark matter halo (dashed red). Recall that the Lagrangian components are defined  at $z=0$ and tracked back in time (see Section \ref{sec_lagrangian}). For most galaxies, the distribution of today's cold gas is more akin to the whole dark matter halo than to the inner dark matter halo, while the opposite is true for today's stars.} 
\label{closeness}
\end{figure}

Fig.~\ref{closeness} shows the distribution of $Q$ values between: cold gas$-$dark matter halo (solid cyan), stars$-$dark matter halo (solid red), cold gas$-$inner dark matter halo (dashed cyan), and stars$-$inner dark matter halo (dashed red). The figure clearly shows that, for most galaxies, today's cold gas follows more closely the specific angular momentum evolution of the whole dark matter halo, while the stars follow the inner dark matter halo. This confirms that the trends hinted at in Fig.~\ref{AM_evolution_examples} are statistically significant and it strongly indicates that the inner dark matter halo 
and the stars lose specific angular momentum through a very similar process (at least for an important fraction of the galaxy population). 

\begin{figure}
\centering
\includegraphics[height=7.2cm,width=8.0cm, trim=0cm 0.5cm 0cm 1.0cm, clip=true]{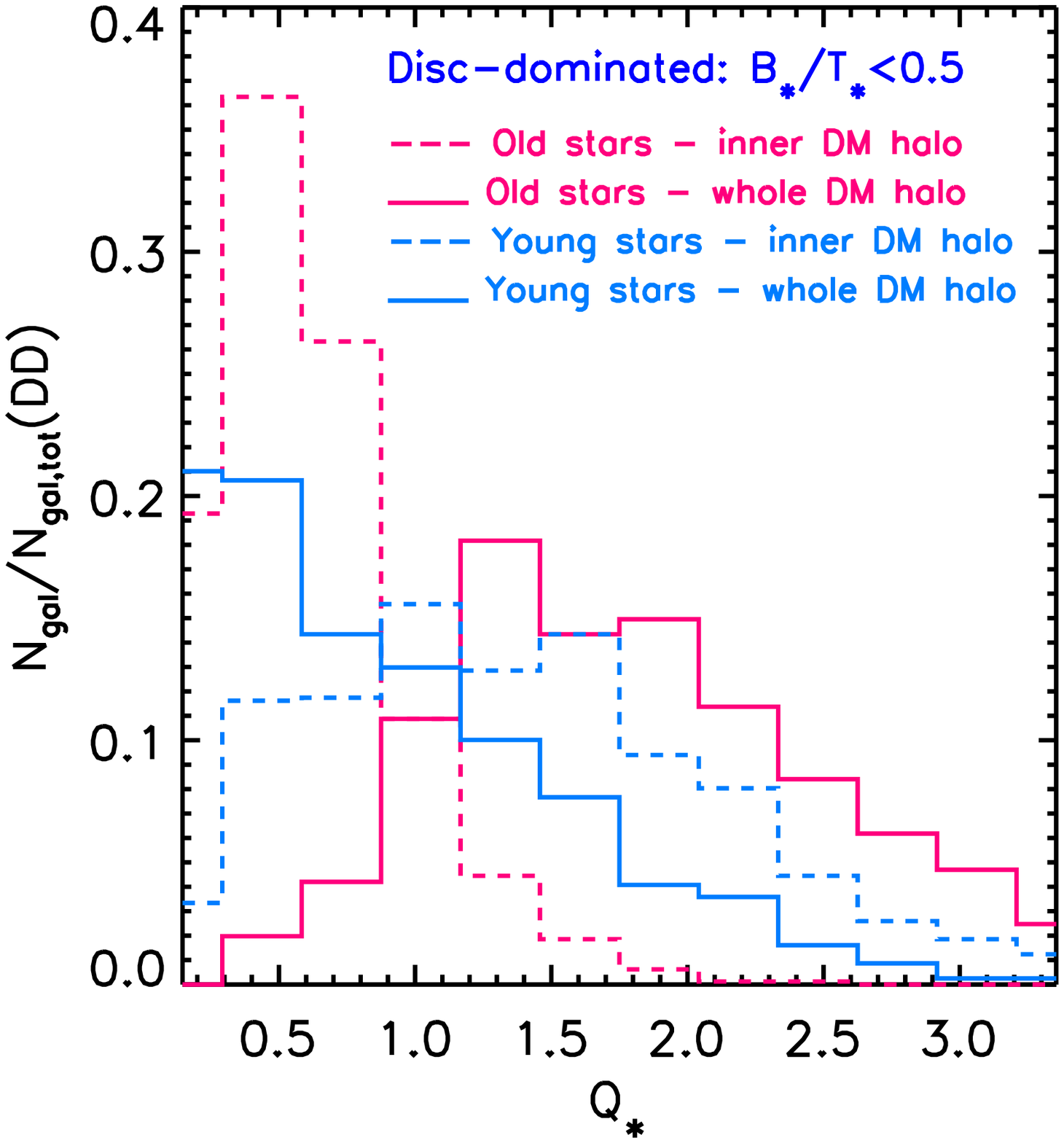} 
\includegraphics[height=7.2cm,width=8.0cm, trim=0cm 0.5cm 0cm 1.0cm, clip=true]{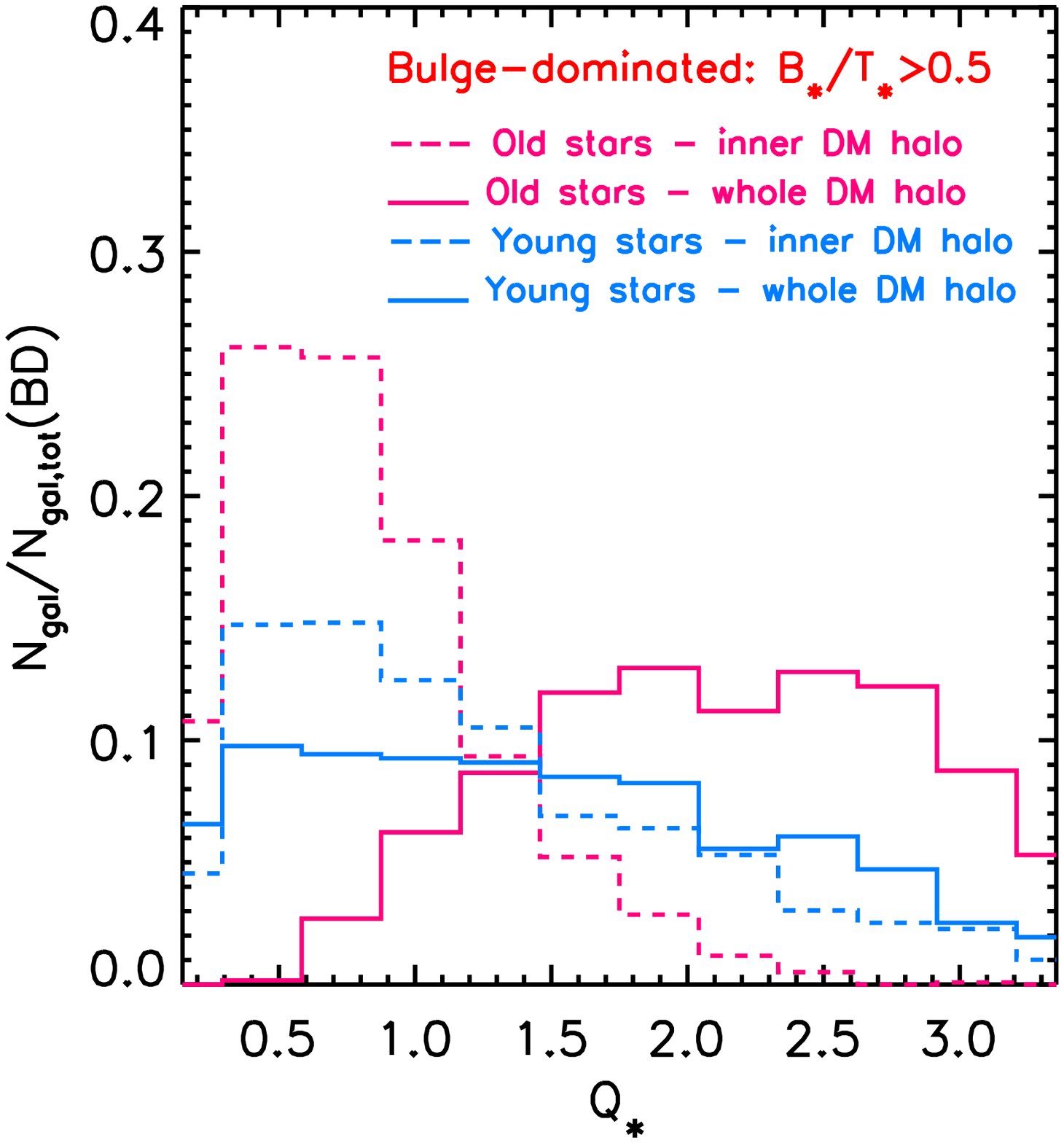} 
\caption{Distribution of the ``closeness" ($Q_\ast$, Eq.~\ref{goodness}) between the specific angular momentum evolution of the Lagrangian components measured after turnaround (of the inner dark matter halo): stars$-$dark matter halo (solid), and stars$-$inner dark matter halo (dashed). The stellar (Lagrangian) component at $z=0$ has been split according to the stellar formation (lookback) time relative to the turnaround epoch: old stars (light red) and young stars (blue). The top (bottom) panel shows those galaxies that are classified as disc-dominated (bulge-dominated) today. Each distribution is normalised to the total number of disc-dominated, $N_{\rm Tot}{\rm (DD)}$, or bulge-dominated, $N_{\rm Tot}{\rm (BD)}$, galaxies in the top and bottom panels, respectively.} 
\label{closeness_age}
\end{figure}

This conclusion is supported by the fact that it is primarily the older stars that most closely trace the specific angular momentum evolution of the inner dark matter halo. This is shown in Fig.~\ref{closeness_age} where we have taken subsamples of the stellar Lagrangian component according to the time when stars were born, relative to the epoch of turnaround of the inner halo $t_{\rm turn}$. Fig.~\ref{closeness_age} is split into two panels according to the morphology of galaxies today: disc-dominated (top panel) and bulge-dominated (bottom panel). The old stars, defined as those born at lookback time $t_L>t_L(\rm turn)$, exhibit a similar distribution of $Q$ values to that of the whole stellar population (Fig.~\ref{closeness}), but their similarity to the inner dark matter halo is stronger (light red dashed lines, particularly for disc-dominated galaxies), while that with the whole dark matter halo is weaker (light red solid lines), particularly for bulge-dominated galaxies. This is expected in a formation scenario where the old stars are already bound to the dark matter subclumps that subsequently form the inner dark matter halo, and therefore share their fate of suffering strong specific angular momentum losses. Fig.~\ref{closeness_age} demonstrates that, in most galaxies, these stars have little affinity with the evolution of the specific angular momentum of the whole dark matter halo, compared to the whole stellar population in Fig.~\ref{closeness}. 
The blue histogram in Fig.~\ref{closeness_age} shows the distribution of young stars with $t_L<t_L(\rm turn)/2$. The situation is quite different from that of the older stars: the affinity with the whole dark matter halo is more evident, particularly for disc-dominated galaxies where the young stars inherit the ``closeness" between the cold gas and the whole dark matter halo. In the case of the bulge-dominated galaxies, this affinity is weaker, which might indicate that an important fraction of the young stars are born from gas that is already bound to the merging dark matter subclumps. 

\begin{figure}
\centering
\includegraphics[height=8.0cm,width=8.0cm]{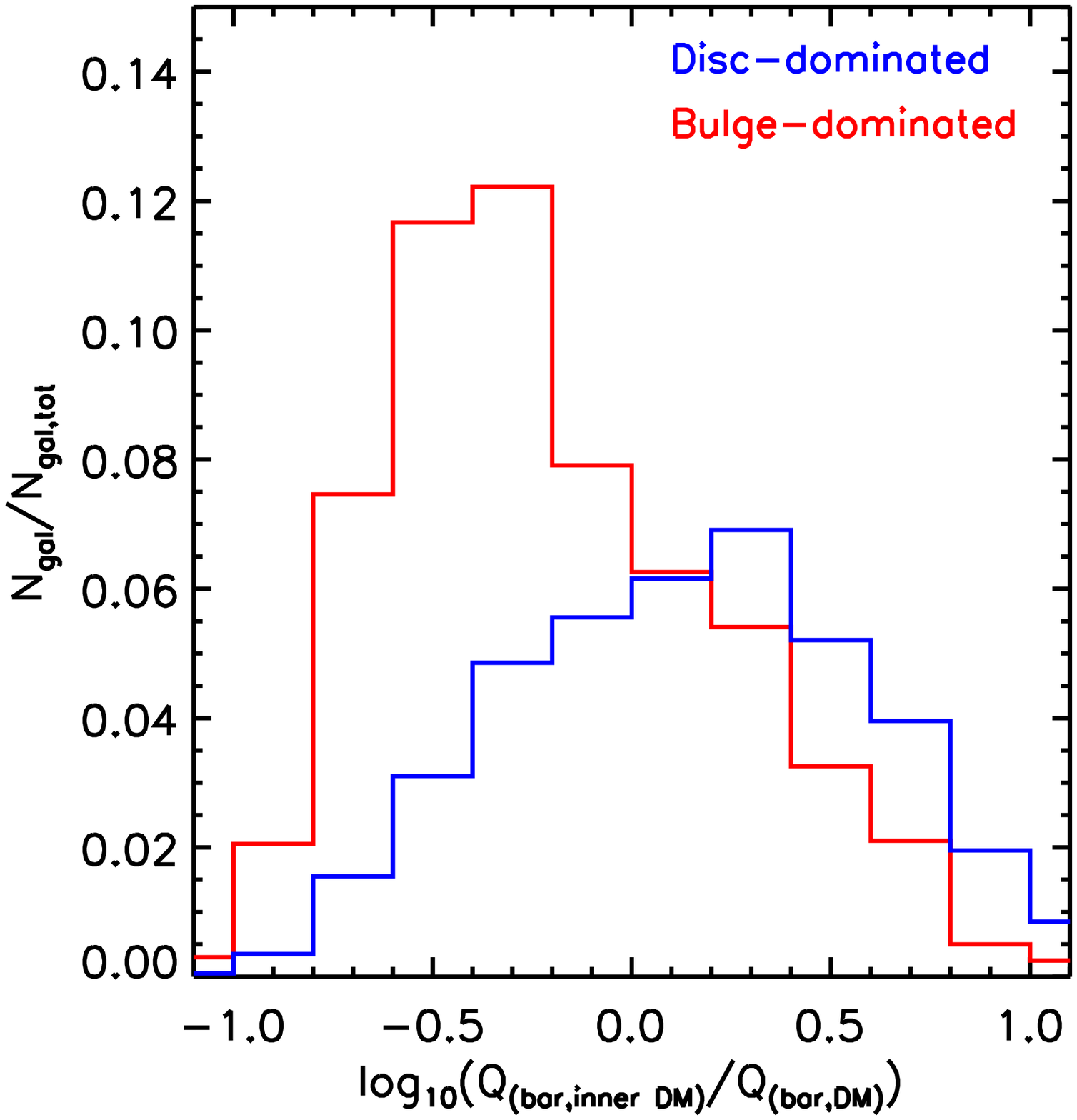} 
\caption{Normalised distribution of the ratio of $Q$ values (Eq.~\ref{goodness}) between the galaxy (stars + cold gas) and the inner dark matter halo $Q_{\rm(bar,inner~DM)}$, and between the galaxy and the whole dark matter halo $Q_{\rm(bar,DM)}$, both measured after turnaround (of the inner dark matter halo). The simulated galaxies have been split into bulge-dominated ($B_\ast/T_\ast\geq0.5$; red) and disc-dominated ($B_\ast/T_\ast<0.5$; blue) according to their circularity distributions today. Each histogram is normalised to the total number of galaxies (both bulge- and disc-dominated). Although there is substantial overlap between the distributions, the evolution of specific angular momentum in most bulge-dominated galaxies is more strongly correlated with that of the inner dark matter halo than with the whole dark matter halo, while the opposite is true for disc-dominated galaxies.} 
\label{Q_ratio}
\end{figure}

Fig.~\ref{closeness_age} shows that the age of today's stars (older/younger relative to the time of turnaround) is an important factor in determining the ``closeness" between the specific angular momentum evolution of the stellar and dark matter components. Nevertheless, it is still instructive to study the stellar population as a whole and attempt to establish a connection between this ``closeness" and the morphology of galaxies today. Fig.~\ref{Q_ratio} shows the distribution of the ratios $Q_{\rm(bar,inner~DM)}/Q_{\rm(bar,DM)}$\footnote{From Eq.~\ref{goodness}, where the subscripts refer to: ``bar'', today's cold baryons (stars + gas), ``inner DM'', inner dark matter halo, and ``DM'', the whole dark matter halo.} for bulge-dominated ($B_\ast/T_\ast\geq0.5$; red) and disc-dominated ($B_\ast/T_\ast<0.5$; blue) galaxies. The bulge-dominated galaxies clearly dominate the population of galaxies where the specific angular momentum of the baryons and of the inner dark matter halo are more closely related. To the right of log($Q_{\rm(bar,inner~DM)}/Q_{\rm(bar,DM)}$)$=0$ in the plot, the situation is less clear, with both samples having similar distributions, although disc-dominated galaxies tend to be more correlated with the whole dark matter halo. The reason why disc-dominated galaxies do not show a stronger ``closeness" to the whole dark matter halo in this plot is that young and old stars have been combined. As is evident from the top panel of Fig.~\ref{closeness_age}, the former are linked to the whole dark matter halo, while the latter are not. In any case, Fig.~\ref{Q_ratio} indicates that a large fraction of today's bulge-dominated galaxies lost their angular momenta in a similar fashion to their host dark matter haloes' inner region.

\section{Specific angular momentum loss and galaxy morphology}\label{sec_AM_morphology}

\begin{figure}
\centering
\includegraphics[height=8.25cm,width=8.25cm]{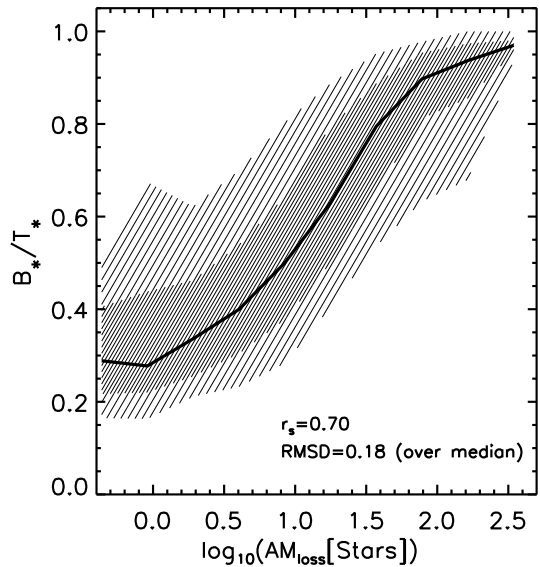} 
\caption{Correlation between the $z=0$ stellar bulge-to-total mass ratio (defined from the circularity distributions) and the maximum loss of specific angular momentum  for the stars in the simulated galaxies. The maximum loss is defined between the time when each component reached its maximum and today (Eq.~\ref{am_loss}). The thick line shows the median of the distribution whereas the hashed areas encompass the $10\%-25\%$, $25\%-75\%$ and $75\%-90\%$ regions. The Spearman's rank correlation coefficient is 0.7 and the root mean square deviation over the median is 0.18.} 
\label{bt_vs_am_loss_baryons}
\end{figure}

In search of a clearer connection between the specific angular momentum evolution of galaxies and their present-day morphology, we show in Fig.~\ref{bt_vs_am_loss_baryons} the stellar bulge-to-total mass ratios for our sample of simulated galaxies at $z=0$ as a function of their relative maximum loss of specific angular momentum:
\begin{equation}\label{am_loss}
	AM_{\rm loss} = \frac{j_{\rm max} - j_0}{j_0},
\end{equation}
where $j_{\rm max}$ is the maximum specific angular momentum of a given component (the stars in the case of Fig.~\ref{bt_vs_am_loss_baryons}) during the evolution, and $j_0$ is its specific angular momentum at $z=0$\footnote{We notice that, because of the way we define the Lagrangian components, at the time where $j_{\rm max}$ is computed, a smaller number of star particles is present compared to those at $z=0$ used to compute $j_0$. This deficit is caused by the fact that some star particles where gas particles at earlier times, which, by definition, do not contribute to the specific angular momentum of the stars. We note that if we take all particle progenitors (stars and gas) of the star particles at $z=0$ to define the specific angular momentum loss of stars in Eq.~\ref{am_loss}, then the main correlations we find in Figs.~\ref{bt_vs_am_loss_baryons} and \ref{money_plot} remain similar but with slightly more scatter.}. 
Fig.~\ref{bt_vs_am_loss_baryons} shows a clear strong correlation (with a Spearman's correlation coefficient $r_s\sim0.7$): the systems with the greatest loss of specific angular momentum end up having the greatest $B_\ast/T_\ast$ values. Since the bulge-to-total mass ratios are computed from the circularity distributions, there is naturally a correlation between $B_\ast/T_\ast$ and $j_0$, but the correlation seen in Fig.~\ref{bt_vs_am_loss_baryons} is not driven purely by this connection. 
To check this, we performed a statistical test using the Akaike Information Criterion (AIC) to compare two models: one where 
$B_\ast/T_\ast$ is described by a polynomial (of degree 3) with only one variable, $j_0$, and one adding an extra variable, 
$j_{\rm max}$. The Residual Sum of Squares of the latter model is smaller than the former by $27\%$, and we find that the latter model is significantly preferred with a bias corrected AIC difference of $\Delta_{AICc}\sim900$ (the relative likelihood between models is exp$(-\Delta_{AICc}/2)$, e.g. see \citealt{AIC2011}).  

Although the correlation in Fig.~\ref{bt_vs_am_loss_baryons} is clear and shows that the specific angular momentum loss and the 
$B_\ast/T_\ast$ value today are linked, 
the dispersion is large. The root mean square deviation, RMSD over the median is 0.18. We find a similar correlation (albeit with more scatter), if we use the baryonic component (stars + today's cold gas) instead of the stars only.

\begin{figure}
\centering
\includegraphics[height=8.25cm,width=8.25cm]{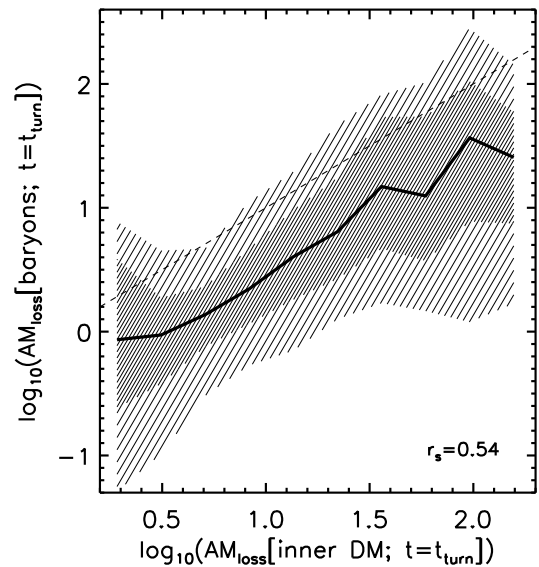} 
\caption{Correlation between the loss of specific angular momentum of the baryons (i.e. stars and cold gas) and the inner dark matter halo (both relative to the time of turnaround of the inner dark matter halo) for the simulated galaxies. The thick line shows the median of the distribution whereas the hashed areas encompass the $10\%-25\%$, $25\%-75\%$ and $75\%-90\%$ regions.
The one-to-one relation is shown as a dashed line. The Spearman's rank correlation coefficient is 0.54.} 
\label{AM_loss_dm10_baryons}
\end{figure}

Fig.~\ref{bt_vs_am_loss_baryons} shows the advantages and limitations of using a single number, $AM_{\rm loss}$, as a ``measure'' of the morphologies of galaxies today, based exclusively on the specific angular momentum at two epochs. It is a very simple quantity that can only pinpoint the $B_\ast/T_\ast$ values with limited accuracy since it cannot fully describe the whole specific angular momentum evolution. 
Since we have shown in previous sections that there is a connection between the evolution of the specific angular momentum of the galaxy and that of the inner dark matter halo (particularly for bulge-dominated galaxies), we now investigate to what extent this connection prevails using a single quantity: the maximum specific angular momentum loss. 

Fig.~\ref{AM_loss_dm10_baryons} shows the correlation between the loss of specific angular momentum of the baryons (i.e. stars and cold gas) and of the inner dark matter halo (both relative to the time of turnaround of the inner dark matter halo) for the simulated galaxies. Although there is a correlation, the dispersion is large with baryons preferentially losing less specific angular momentum than the inner dark matter halo (as illustrated by the reference one-to-one dashed line). 
The dispersion is a reflection of the deviations between the detailed evolution of specific angular momentum of both components: the residuals of the one-to-one relation anti-correlate strongly ($r_s=-0.66$) with the ``closeness" ratio $Q_{\rm (bar,inner~DM)}/Q_{\rm (bar,DM)}$, i.e., the larger the deviation from the one-to-one relation, the larger the difference between the evolutionary tracks of the cold baryons and the inner dark matter halo.
This indicates that the loss of specific angular momentum in the inner dark matter halo and in the baryons are only correlated 
for a subsample of the galaxy population. This is something already hinted at by the results of Section \ref{sec_correspond} (see Fig.~\ref{closeness}). 

Remarkably, and despite this complication, there is a reasonably tight correlation between the {\it maximum} loss of specific angular momentum of the stars and the inner dark matter halo (see Eq.~\ref{am_loss}), which is shown in Fig.~\ref{money_plot}. 
 The thick black line shows the median of the distribution whereas the hashed black areas encompass the $10\%-25\%$, $25\%-75\%$ and $75\%-90\%$ regions. Notice that the median is very close to the one-to-one relation (dashed line). The correlation is even stronger if we consider only those galaxies for which the specific angular momentum history of the stars and the inner dark matter halo track each other closely ($Q_{\rm(\ast,inner~DM)}<0.7$, green lines encompassing $10\%-90\%$ of the distribution in this reduced sample). 
 Still, the dispersion remains large (the $RMSD$ from the one-to-one relation is $\sim0.35$).  Below we comment on some of the physical quantities behind this dispersion. We notice that bulge-dominated galaxies (having lost most of their angular momenta) are located mostly in the upper-right of the distribution (as can be seen in the projected distribution in the Y axis shown by the red histogram to the right of Fig.~\ref{money_plot}), while disc-dominated galaxies populate the lower-left (shown by the blue histogram). Combining Figs.~\ref{money_plot} and  \ref{bt_vs_am_loss_baryons}, we have a way to connect (statistically) the stellar bulge-to-total mass ratios (at $z=0$) and the loss of specific angular momentum of the inner dark matter halo. 

\begin{figure}
\centering
\includegraphics[height=8.0cm,width=8.75cm, trim=0cm 0.5cm 0cm 0cm, clip=true]{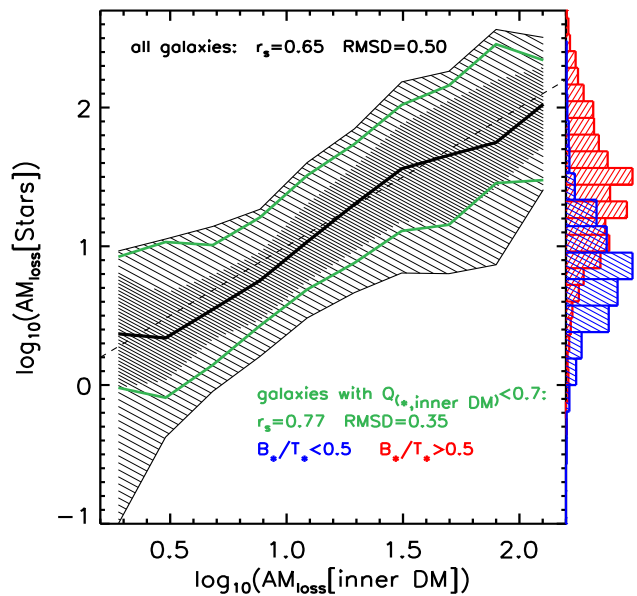} 
\caption{Correlation between the maximum loss of specific angular momentum of the stars and of the inner dark matter halo for the simulated galaxies. The thick black line shows the median of the distribution whereas the hashed black areas encompass the $10\%-25\%$, $25\%-75\%$ and $75\%-90\%$ regions. The $10\%-90\%$ distribution of those galaxies for which the specific angular momentum evolution of the stars and of the inner dark matter halo is close ($Q_{\rm (\ast,inner~DM)}<0.7$) is bracketed by the solid green lines. The Spearman's rank correlation coefficients and the (logarithmic) root mean square standard deviation are shown in the legend, for each case. The green subsample is further divided according to the morphology of the galaxies today, disc- and bulge-dominated. Their respective projected distributions in the Y axis ($AM_{\rm loss}(\rm stars)$) are shown in blue and red to the right of the plot. } 
\label{money_plot}
\end{figure}


Regarding the issue of limited sampling of stars for the smallest galaxies in the simulation we use, and its connection with the physical resolution limit (see discussion at the end of Section \ref{sec_morpho}), we did not find a systematic effect in our main result if we discard the smallest galaxies ($M_\ast<7.5\times10^{9}$~M$_\odot$). That is, the correspondence between the specific angular momentum loss of the inner dark matter halo and that of the stellar component shown in Fig.~\ref{money_plot} is also valid for these galaxies
The changes to the Spearman's rank correlation coefficient and the $RMSD$ are at the percent level when we discard the smallest galaxies. We also applied a Kolmogorov-Smirnov (K-S) test to the unbinned $AM_{\rm loss}(\rm stars)$ distributions of two subsamples of the low-mass galaxies split by their stellar mass. The first sample with $M_\ast<7.5\times10^{9}$~M$_\odot$ and the second sample with $7.5\times10^{9}$M$_\odot<M_\ast<1.5\times10^{10}$M$_\odot$, with 468 and 529 galaxies, respectively. We found a K-S statistic of $D=0.06$ with a probability of $0.24$. Repeating the same test but for $AM_{\rm loss}(\rm inner~DM)$, we found $D=0.03$ with a probability of $0.89$. The test is inconclusive for the stars while for the inner dark matter halo, the
high probability indicates that the low-mass and high-mass subsamples likely originate from similar distributions. The lower probability in the case of the stars however, might indicate an imprint of the physical resolution limit since the collisionless dark matter is not subject to the subgrid pressure of the imposed temperature floor in EAGLE, which indirectly impacts the stars since they are formed from the gas. One must remember however, that there is also a physical effect associated with the stronger prominence of cold gas at lower stellar masses, from which the 
stellar component would then tend to inherit higher specific angular momentum. This creates a slight mass dependence in the median of the $AM_{\rm loss}(\rm stars)$ distribution. This physical effect also impacts the interpretation of the K-S test. 

Nevertheless, as stated before, our main result (Fig.~\ref{money_plot}) is not significantly altered by limited sampling. On the other hand, this seems to be an issue in the correlation between the $B_\ast/T_\ast(z=0)$ and $AM_{\rm loss}(\rm stars)$ shown in Fig.~\ref{bt_vs_am_loss_baryons}. The galaxies with $M_\ast<7.5\times10^{9}$~M$_\odot$ would appear uncorrelated in this plot, which seems to verify the concerns regarding the physical resolution in the simulation we analysed, and expressed at the end of Section \ref{sec_morpho}. In particular, the stellar bulge-to-total mass ratios seem to be strongly affected by this issue, which is the reason why we decided to exclude galaxies below this stellar mass threshold from most of our results. 

\subsection{Additional drivers of specific angular momentum loss}

We have searched thoroughly for a third parameter that reduces the scatter in Fig.~\ref{money_plot} but we have not been able to find one among the dark matter quantities that is statistically significant. We explored the formation time of the inner dark matter halo, the specific angular momentum loss of the whole dark matter halo, and the misalignment of the specific angular momentum of the inner dark matter between turnaround and today. The (logarithmic) Residual Sum of Squares ($RSS_{\rm log}$, from the best linear fit) is reduced by less than $3\%$ in all cases, and since we are adding one variable to the multivariate regression fit, this reduction is not significant.

Instead, we have found that the scatter in  Fig.~\ref{money_plot} correlates more strongly with the properties of today's cold baryons. In particular, the deficit between the total mass of the Lagrangian gas component that is transformed into stars between turnaround and today, $\Delta M_{\rm gas}=M_{\rm gas}(z_{\rm turn})-M_{\rm gas}(z=0)$, and the net mass gain of the stellar component during this time, $\Delta M_{\ast}=M_\ast(z=0)-M_\ast(z_{\rm turn})$. The ratio $\Delta M_{\ast}/\Delta M_{\rm gas}$ anticorrelates with the residuals of the correlation in Fig.~\ref{money_plot} with $r_s\sim-0.43$. This mass deficit is partially related to the baryons that are lost during stellar feedback (i.e. the mass loss in the simulated stellar particles during stellar and supernovae winds), $\Delta M_{\ast}/\Delta M_{\rm gas}<1$, and are not present in the Lagrangian components at $z=0$. The larger the deficit, the stronger the loss of specific angular momentum of the stellar component: galaxies below (above) the one-to-one relation in Fig.~\ref{money_plot} have a mean and standard deviation for $\Delta M_{\ast}/\Delta M_{\rm gas}: 0.72\pm0.1~(0.63\pm0.27)$. This can be interpreted as follows. During an episode of stellar mass loss caused by a supernova-driven galactic wind, energy and momentum are deposited into the immediate neighborhood of the star. In particular, the gas particles around the star particle are pushed by the pressure gradient, which eventually have a dynamical (gravitational) impact on other nearby star particles. Several episodes like this tend to, on average, isotropise the orbits of the star particles, causing a net loss of the total specific angular momentum of the stellar component. This also impacts the dark matter particles, but to a lesser extent since they tend to have (near the halo centre) more isotropic orbits than the stars to begin with. This effect is more noticeable when this stellar mass loss (and subsequent loss of specific angular momentum) is not compensated by new stars being born from gas with high specific angular momentum, i.e. when the stellar mass deficit  $\Delta M_{\ast}/\Delta M_{\rm gas}$ is high.

Of similar significance for the scatter seen in Fig.~\ref{money_plot}, is the relative loss of specific angular momentum in today's cold gas (see Eq.~\ref{am_loss}). This variable 
correlates with the residuals of the correlation in Fig.~\ref{money_plot} with $r_s\sim0.40$ (adding this as a third parameter reduces $RSS_{\rm log}$ from the best linear fit by $16\%$). 
This is clearly a consequence of the young stars inheriting
the high specific angular momentum of the gas they are born from. The lower the loss of specific angular momentum of the gas, the more likely it is that the stellar component that is being fed by this gas will have lower specific angular momentum losses.



\section{Discussion and Conclusions} \label{sec_Conclusions}

Using a state-of-the-art hydrodynamical simulation from the EAGLE project \citep[labeled Ref-L100N1504,][]{Schaye2015,Crain2015}, we have investigated the connection between today's galaxy morphologies and the evolution of the specific angular momentum (in physical units) of the Lagrangian components that constitute the galaxy today. These components are the dark matter halo (with the virial radius as a boundary), the inner dark matter halo (within $10\%$ of the virial radius) and the cold baryons (stars and radiatively cooled dense gas). See Section \ref{sec_lagrangian} for a description of these components and how they are traced back in time. In particular, our analysis follows the methodology of \citet{Zavala2008} in defining the galaxy at $z=0$ and tracing back in time its progenitor particles, rather than defining the galaxy independently at each time. In this way, we can follow the history of
gain/loss of specific angular momentum of the matter that comprise a galaxy today.

We use a sample of nearly 2500 simulated central galaxies from the EAGLE project with $M_{\ast}(z=0)\ge7.5\times10^9$M$_\odot$, and classified them according to their stellar bulge-to-total mass ratios $B_\ast/T_\ast$, computed using the circularity distribution of their stars today (see Section \ref{sec_morpho}). From this sample\footnote{We note that, in order to avoid poor particle sampling, our results from Fig.~\ref{am_st_vs_dm10} onwards are actually based on a smaller subsample of 2005 galaxies, which have at least 500 particles at the time of turnaround of the inner dark matter halo (see Section~\ref{sec_AM_evolution}, particularly the text related to Fig.~\ref{am_st_vs_dm10}). On the other hand, the results shown in Fig.~\ref{AM_Fall} were obtained from our original larger sample of 4000 galaxies with $M_\ast(z=0)\ge4\times10^8$M$_\odot$, in order to show the low-mass end of this plot.}, we draw the following main conclusions:
\begin{itemize}
	\item Our kinematical classification splits the galaxies in two types, bulge-dominated ($B_\ast/T_\ast\geq0.5$) and disc-dominated ($B_\ast/T_\ast<0.5$), which are separated in the $j_\ast-M_\ast$ plane in a way that is roughly consistent with what is observed (see Fig.~\ref{AM_Fall}). 
	\item Prior to the epoch during which the inner dark matter protohalo reaches its maximum expansion and begins to collapse (turnaround), all galactic components follow the expectations from the
	spherical collapse model and the classical tidal torque theory. They occupy a similar Lagrangian region and acquire angular momentum from environmental torques at the same rate. They reach their maximum specific angular momentum around this turnaround epoch (see Fig.~\ref{AM_evolution}).
	\item After turnaround, the specific angular momentum evolution of the different components diverges: whereas the whole dark matter halo conserves most of its specific angular momentum, the inner dark matter halo loses $\sim90\%$ (see Fig.~\ref{AM_evolution}), presumably as 
	a consequence of angular momentum transfer to the outer halo during its assembly (dominated by mergers of dark matter subclumps). This confirms the results first noted by \citet{Frenk1985} and explored in detail in 	   \citet{Zavala2008}. 
	\item On the other hand, the morphology of galaxies today is strongly correlated with the loss of specific angular momentum after turnaround (see Fig.~\ref{bt_vs_am_loss_baryons}). Bulge-dominated galaxies, which we define as those with $B_\ast/T_\ast>0.5$, have lost $80\%$ (median) of their maximum specific angular momentum (attained near turnaround), while for disc-dominated galaxies ($B_\ast/T_\ast<0.5$), the median loss is $50\%$ (see Fig.~\ref{AM_evolution}).
	\item We find remarkable links between the specific angular momentum evolution of the dark and baryonic components. We find that the specific angular momentum history of the cold gas (stars) of today's galaxies is statistically strongly correlated with the history of the whole (inner) dark matter halo (see Fig.~\ref{closeness}, and also Fig.~\ref{AM_evolution_examples} for examples that illustrate this behaviour). The connection between the stellar and dark components depends on the ages of the stars: the older stars (those formed before turnaround) track closely the specific angular momentum of the inner dark matter halo, while the younger stars (those formed after turnaround) are indirectly correlated with the whole dark matter halo since they are born from recently accreted gas, which carries high specific angular momentum and has an affinity with the whole dark matter halo (see Fig.~\ref{closeness_age}).
	\item We therefore obtain the striking result that the assembly of the inner dark matter halo and its history of specific angular momentum loss is linked to the morphology of galaxies today. In particular, we have found that there is a nearly one-to-one correlation between the loss of specific angular momentum of the stellar component and that of the inner dark matter halo (see Fig.~\ref{money_plot}). This agrees well with the finding that the distribution of stars is much better aligned with the inner dark matter halo than with the whole dark matter halo \citep{Vell2015}. Although the dispersion in this correlation is large, the result implies that we can infer (statistically) the assembly history of the inner dark matter halo indirectly through the morphology of the galaxy it hosts (via the loss of specific angular momentum of the stars).	
\end{itemize}

The latter is our most significant result, which highlights that in a physical model of the formation and evolution of galaxies (as represented by the reference EAGLE simulation we use), the histories of star formation efficiency in the galaxy and assembly of the inner dark matter halo, work in synergy to create a common evolution of the specific angular momentum of the stars and of the inner dark matter halo. The morphology of galaxies today is intimately connected to this common evolution.
We notice also that the correlation in Fig.~\ref{money_plot} is, to our knowledge, the strongest link reported between galaxy morphology (albeit indirectly through Fig.~\ref{bt_vs_am_loss_baryons}) and a global property of dark matter haloes in recent hydrodynamical simulations. For instance, \citet{Sales2012} made a careful exploration (using the GIMIC simulations \citealt{Crain2009}) of different correlations between galaxy morphology and diverse global properties. One of their conclusions was that the morphological features of galaxies at $z=0$ were very poorly correlated with their halo properties. Although statistically weak, they did however find that spheroids today tend to have stronger misalignments than discs between the specific angular momentum measured at the time of turnaround and today; particularly when the misalignment is measured in the inner regions of the halo at these different times (see their Fig. 8). They then suggested that ``...it might be possible to use the angular momentum properties of a dark matter halo at turnaround to `predict' the morphology of its central galaxy at $z = 0$...". We agree with this conclusion, but instead of basing it on the coherent alignment of angular momentum during the assembly of the galaxy, we derive it directly from a
strong correlation connecting the specific angular momentum loss of the stars and the inner dark matter halo between turnaround and today. We emphasize that finding this correlation was only possible by the use of a methodology that tracks the evolution of the Lagrangian region that defines the galactic system today. 



\section*{Acknowledgments}

The Dark Cosmology Centre is funded by the DNRF. JZ is supported by
the EU under a Marie Curie International
Incoming Fellowship, contract PIIF-GA-2013-62772. This work was supported by the Science and Technology Facilities Council (grant number ST/F001166/1); 
European Research 
Council (grant numbers GA 267291 ``Cosmiway'' and GA 278594 
``GasAroundGalaxies'') and by 
the Interuniversity Attraction Poles Programme 
initiated by the Belgian Science Policy Office (AP P7/08 CHARM). 
RAC is a Royal 
Society University Research Fellow.  This work used the DiRAC Data Centric 
system at Durham 
University, operated by the Institute for Computational 
Cosmology on behalf of the STFC DiRAC HPC Facility 
(www.dirac.ac.uk). This 
equipment was funded by BIS National E-infrastructure capital grant
 ST/K00042X/1, STFC capital 
grant ST/H008519/1, and STFC DiRAC Operations grant 
ST/K003267/1 and Durham University. DiRAC is part of the National 
 E-Infrastructure.  We acknowledge PRACE for awarding us access to the Curie 
machine based in France at TGCC, CEA, Bruy\`eres-le-Ch\^atel.

\bibliography{lit}

\begin{thebibliography}{}
\makeatletter
\relax
\def\mn@urlcharsother{\let\do\@makeother \do\$\do\&\do\#\do\^\do\_\do\%\do\~}
\def\mn@doi{\begingroup\mn@urlcharsother \@ifnextchar [ {\mn@doi@}
  {\mn@doi@[]}}
\def\mn@doi@[#1]#2{\def\@tempa{#1}\ifx\@tempa\@empty \href
  {http://dx.doi.org/#2} {doi:#2}\else \href {http://dx.doi.org/#2} {#1}\fi
  \endgroup}
\def\mn@eprint#1#2{\mn@eprint@#1:#2::\@nil}
\def\mn@eprint@arXiv#1{\href {http://arxiv.org/abs/#1} {{\tt arXiv:#1}}}
\def\mn@eprint@dblp#1{\href {http://dblp.uni-trier.de/rec/bibtex/#1.xml}
  {dblp:#1}}
\def\mn@eprint@#1:#2:#3:#4\@nil{\def\@tempa {#1}\def\@tempb {#2}\def\@tempc
  {#3}\ifx \@tempc \@empty \let \@tempc \@tempb \let \@tempb \@tempa \fi \ifx
  \@tempb \@empty \def\@tempb {arXiv}\fi \@ifundefined
  {mn@eprint@\@tempb}{\@tempb:\@tempc}{\expandafter \expandafter \csname
  mn@eprint@\@tempb\endcsname \expandafter{\@tempc}}}

\bibitem[\protect\citeauthoryear{{Abadi}, {Navarro}, {Steinmetz}  \&
  {Eke}}{{Abadi} et~al.}{2003}]{Abadi2003}
{Abadi} M.~G.,  {Navarro} J.~F.,  {Steinmetz} M.,   {Eke} V.~R.,  2003, \mn@doi
  [\apj] {10.1086/378316}, \href
  {http://adsabs.harvard.edu/abs/2003ApJ...597...21A} {597, 21}

\bibitem[\protect\citeauthoryear{{Bailin} et~al.,}{{Bailin}
  et~al.}{2005}]{Bailin2005}
{Bailin} J.,  et~al., 2005, \mn@doi [\apjl] {10.1086/432157}, \href
  {http://adsabs.harvard.edu/abs/2005ApJ...627L..17B} {627, L17}

\bibitem[\protect\citeauthoryear{{Bett}, {Eke}, {Frenk}, {Jenkins}, {Helly}  \&
  {Navarro}}{{Bett} et~al.}{2007}]{Bett2007}
{Bett} P.,  {Eke} V.,  {Frenk} C.~S.,  {Jenkins} A.,  {Helly} J.,   {Navarro}
  J.,  2007, \mn@doi [\mnras] {10.1111/j.1365-2966.2007.11432.x}, \href
  {http://adsabs.harvard.edu/abs/2007MNRAS.376..215B} {376, 215}

\bibitem[\protect\citeauthoryear{{Bett}, {Eke}, {Frenk}, {Jenkins}  \&
  {Okamoto}}{{Bett} et~al.}{2010}]{Bett2010}
{Bett} P.,  {Eke} V.,  {Frenk} C.~S.,  {Jenkins} A.,   {Okamoto} T.,  2010,
  \mn@doi [\mnras] {10.1111/j.1365-2966.2010.16368.x}, \href
  {http://adsabs.harvard.edu/abs/2010MNRAS.404.1137B} {404, 1137}

\bibitem[\protect\citeauthoryear{{Bower}, {Benson}, {Malbon}, {Helly}, {Frenk},
  {Baugh}, {Cole}  \& {Lacey}}{{Bower} et~al.}{2006}]{Bower2006}
{Bower} R.~G.,  {Benson} A.~J.,  {Malbon} R.,  {Helly} J.~C.,  {Frenk} C.~S.,
  {Baugh} C.~M.,  {Cole} S.,   {Lacey} C.~G.,  2006, \mn@doi [\mnras]
  {10.1111/j.1365-2966.2006.10519.x}, \href
  {http://adsabs.harvard.edu/abs/2006MNRAS.370..645B} {370, 645}

\bibitem[\protect\citeauthoryear{{Brook} et~al.,}{{Brook}
  et~al.}{2011}]{Brook2011}
{Brook} C.~B.,  et~al., 2011, \mn@doi [\mnras]
  {10.1111/j.1365-2966.2011.18545.x}, \href
  {http://adsabs.harvard.edu/abs/2011MNRAS.415.1051B} {415, 1051}

\bibitem[\protect\citeauthoryear{{Camps} \& {Baes}}{{Camps} \&
  {Baes}}{2015}]{Camps2015}
{Camps} P.,  {Baes} M.,  2015, \mn@doi [Astronomy and Computing]
  {10.1016/j.ascom.2014.10.004}, \href
  {http://adsabs.harvard.edu/abs/2015A%26C.....9...20C} {9, 20}

\bibitem[\protect\citeauthoryear{{Catelan} \& {Theuns}}{{Catelan} \&
  {Theuns}}{1996a}]{Catelan1996a}
{Catelan} P.,  {Theuns} T.,  1996a, \mnras, \href
  {http://adsabs.harvard.edu/abs/1996MNRAS.282..436C} {282, 436}

\bibitem[\protect\citeauthoryear{{Catelan} \& {Theuns}}{{Catelan} \&
  {Theuns}}{1996b}]{Catelan1996b}
{Catelan} P.,  {Theuns} T.,  1996b, \mnras, \href
  {http://adsabs.harvard.edu/abs/1996MNRAS.282..455C} {282, 455}

\bibitem[\protect\citeauthoryear{{Crain} et~al.,}{{Crain}
  et~al.}{2009}]{Crain2009}
{Crain} R.~A.,  et~al., 2009, \mn@doi [\mnras]
  {10.1111/j.1365-2966.2009.15402.x}, \href
  {http://adsabs.harvard.edu/abs/2009MNRAS.399.1773C} {399, 1773}

\bibitem[\protect\citeauthoryear{{Crain} et~al.,}{{Crain}
  et~al.}{2015}]{Crain2015}
{Crain} R.~A.,  et~al., 2015, \mn@doi [\mnras] {10.1093/mnras/stv725}, \href
  {http://adsabs.harvard.edu/abs/2015MNRAS.450.1937C} {450, 1937}

\bibitem[\protect\citeauthoryear{{Croton} et~al.,}{{Croton}
  et~al.}{2006}]{Croton2006}
{Croton} D.~J.,  et~al., 2006, \mn@doi [\mnras]
  {10.1111/j.1365-2966.2005.09675.x}, \href
  {http://adsabs.harvard.edu/abs/2006MNRAS.365...11C} {365, 11}

\bibitem[\protect\citeauthoryear{{D'Onghia} \& {Navarro}}{{D'Onghia} \&
  {Navarro}}{2007}]{Donghia2007}
{D'Onghia} E.,  {Navarro} J.~F.,  2007, \mn@doi [\mnras]
  {10.1111/j.1745-3933.2007.00348.x}, \href
  {http://adsabs.harvard.edu/abs/2007MNRAS.380L..58D} {380, L58}

\bibitem[\protect\citeauthoryear{{Dalla Vecchia} \& {Schaye}}{{Dalla Vecchia}
  \& {Schaye}}{2012}]{DV2012}
{Dalla Vecchia} C.,  {Schaye} J.,  2012, \mn@doi [\mnras]
  {10.1111/j.1365-2966.2012.21704.x}, \href
  {http://adsabs.harvard.edu/abs/2012MNRAS.426..140D} {426, 140}

\bibitem[\protect\citeauthoryear{{Danovich}, {Dekel}, {Hahn}, {Ceverino}  \&
  {Primack}}{{Danovich} et~al.}{2015}]{Danovich2015}
{Danovich} M.,  {Dekel} A.,  {Hahn} O.,  {Ceverino} D.,   {Primack} J.,  2015,
  \mn@doi [\mnras] {10.1093/mnras/stv270}, \href
  {http://adsabs.harvard.edu/abs/2015MNRAS.449.2087D} {449, 2087}

\bibitem[\protect\citeauthoryear{{Dekel} et~al.,}{{Dekel}
  et~al.}{2009}]{Dekel2009}
{Dekel} A.,  et~al., 2009, \mn@doi [\nat] {10.1038/nature07648}, \href
  {http://adsabs.harvard.edu/abs/2009Natur.457..451D} {457, 451}

\bibitem[\protect\citeauthoryear{{Dolag}, {Borgani}, {Murante}  \&
  {Springel}}{{Dolag} et~al.}{2009}]{Dolag2009}
{Dolag} K.,  {Borgani} S.,  {Murante} G.,   {Springel} V.,  2009, \mn@doi
  [\mnras] {10.1111/j.1365-2966.2009.15034.x}, \href
  {http://adsabs.harvard.edu/abs/2009MNRAS.399..497D} {399, 497}

\bibitem[\protect\citeauthoryear{{Doroshkevich}}{{Doroshkevich}}{1970}]{Doro1970}
{Doroshkevich} A.~G.,  1970, Astrofizika, \href
  {http://adsabs.harvard.edu/abs/1970Afz.....6..581D} {6, 581}

\bibitem[\protect\citeauthoryear{{Dubois} et~al.,}{{Dubois}
  et~al.}{2014}]{Dubois2014}
{Dubois} Y.,  et~al., 2014, \mn@doi [\mnras] {10.1093/mnras/stu1227}, \href
  {http://adsabs.harvard.edu/abs/2014MNRAS.444.1453D} {444, 1453}

\bibitem[\protect\citeauthoryear{{Durier} \& {Dalla Vecchia}}{{Durier} \&
  {Dalla Vecchia}}{2012}]{Durier2012}
{Durier} F.,  {Dalla Vecchia} C.,  2012, \mn@doi [\mnras]
  {10.1111/j.1365-2966.2011.19712.x}, \href
  {http://adsabs.harvard.edu/abs/2012MNRAS.419..465D} {419, 465}

\bibitem[\protect\citeauthoryear{{Fall}}{{Fall}}{1983}]{Fall1983}
{Fall} S.~M.,  1983, in {Athanassoula} E.,  ed.,  IAU Symposium Vol. 100,
  Internal Kinematics and Dynamics of Galaxies. pp 391--398

\bibitem[\protect\citeauthoryear{{Fall} \& {Efstathiou}}{{Fall} \&
  {Efstathiou}}{1980}]{Fall1980}
{Fall} S.~M.,  {Efstathiou} G.,  1980, \mnras, \href
  {http://adsabs.harvard.edu/abs/1980MNRAS.193..189F} {193, 189}

\bibitem[\protect\citeauthoryear{{Fall} \& {Romanowsky}}{{Fall} \&
  {Romanowsky}}{2013}]{Fall2013}
{Fall} S.~M.,  {Romanowsky} A.~J.,  2013, \mn@doi [\apjl]
  {10.1088/2041-8205/769/2/L26}, \href
  {http://adsabs.harvard.edu/abs/2013ApJ...769L..26F} {769, L26}

\bibitem[\protect\citeauthoryear{{Fardal}, {Katz}, {Gardner}, {Hernquist},
  {Weinberg}  \& {Dav{\'e}}}{{Fardal} et~al.}{2001}]{Fardal2001}
{Fardal} M.~A.,  {Katz} N.,  {Gardner} J.~P.,  {Hernquist} L.,  {Weinberg}
  D.~H.,   {Dav{\'e}} R.,  2001, \mn@doi [\apj] {10.1086/323519}, \href
  {http://adsabs.harvard.edu/abs/2001ApJ...562..605F} {562, 605}

\bibitem[\protect\citeauthoryear{{Fisher} \& {Drory}}{{Fisher} \&
  {Drory}}{2011}]{Fisher2011}
{Fisher} D.~B.,  {Drory} N.,  2011, \mn@doi [\apjl]
  {10.1088/2041-8205/733/2/L47}, \href
  {http://adsabs.harvard.edu/abs/2011ApJ...733L..47F} {733, L47}

\bibitem[\protect\citeauthoryear{{Frenk}, {White}, {Efstathiou}  \&
  {Davis}}{{Frenk} et~al.}{1985}]{Frenk1985}
{Frenk} C.~S.,  {White} S.~D.~M.,  {Efstathiou} G.,   {Davis} M.,  1985,
  \mn@doi [\nat] {10.1038/317595a0}, \href
  {http://adsabs.harvard.edu/abs/1985Natur.317..595F} {317, 595}

\bibitem[\protect\citeauthoryear{{Furlong} et~al.,}{{Furlong}
  et~al.}{2015}]{Furlong2015}
{Furlong} M.,  et~al., 2015, \mn@doi [\mnras] {10.1093/mnras/stv852}, \href
  {http://adsabs.harvard.edu/abs/2015MNRAS.450.4486F} {450, 4486}

\bibitem[\protect\citeauthoryear{{Gadotti}}{{Gadotti}}{2009}]{Gadotti2009}
{Gadotti} D.~A.,  2009, \mn@doi [\mnras] {10.1111/j.1365-2966.2008.14257.x},
  \href {http://adsabs.harvard.edu/abs/2009MNRAS.393.1531G} {393, 1531}

\bibitem[\protect\citeauthoryear{{Genel}, {Fall}, {Hernquist}, {Vogelsberger},
  {Snyder}, {Rodriguez-Gomez}, {Sijacki}  \& {Springel}}{{Genel}
  et~al.}{2015}]{Genel2015}
{Genel} S.,  {Fall} S.~M.,  {Hernquist} L.,  {Vogelsberger} M.,  {Snyder}
  G.~F.,  {Rodriguez-Gomez} V.,  {Sijacki} D.,   {Springel} V.,  2015, \mn@doi
  [\apjl] {10.1088/2041-8205/804/2/L40}, \href
  {http://adsabs.harvard.edu/abs/2015ApJ...804L..40G} {804, L40}

\bibitem[\protect\citeauthoryear{{Governato} et~al.,}{{Governato}
  et~al.}{2009}]{Governato2009}
{Governato} F.,  et~al., 2009, \mn@doi [\mnras]
  {10.1111/j.1365-2966.2009.15143.x}, \href
  {http://adsabs.harvard.edu/abs/2009MNRAS.398..312G} {398, 312}

\bibitem[\protect\citeauthoryear{{Hopkins}}{{Hopkins}}{2013}]{Hopkins2013}
{Hopkins} P.~F.,  2013, \mn@doi [\mnras] {10.1093/mnras/sts210}, \href
  {http://adsabs.harvard.edu/abs/2013MNRAS.428.2840H} {428, 2840}

\bibitem[\protect\citeauthoryear{{Hopkins} et~al.,}{{Hopkins}
  et~al.}{2009}]{Hopkins2009}
{Hopkins} P.~F.,  et~al., 2009, \mn@doi [\mnras]
  {10.1111/j.1365-2966.2009.14983.x}, \href
  {http://adsabs.harvard.edu/abs/2009MNRAS.397..802H} {397, 802}

\bibitem[\protect\citeauthoryear{{Hopkins} et~al.,}{{Hopkins}
  et~al.}{2010}]{Hopkins2010}
{Hopkins} P.~F.,  et~al., 2010, \mn@doi [\apj] {10.1088/0004-637X/715/1/202},
  \href {http://adsabs.harvard.edu/abs/2010ApJ...715..202H} {715, 202}

\bibitem[\protect\citeauthoryear{{Kauffmann}, {White}  \&
  {Guiderdoni}}{{Kauffmann} et~al.}{1993}]{Kauffmann1993}
{Kauffmann} G.,  {White} S.~D.~M.,   {Guiderdoni} B.,  1993, \mnras, \href
  {http://adsabs.harvard.edu/abs/1993MNRAS.264..201K} {264, 201}

\bibitem[\protect\citeauthoryear{{Kere{\v s}}, {Katz}, {Weinberg}  \&
  {Dav{\'e}}}{{Kere{\v s}} et~al.}{2005}]{Keres2005}
{Kere{\v s}} D.,  {Katz} N.,  {Weinberg} D.~H.,   {Dav{\'e}} R.,  2005, \mn@doi
  [\mnras] {10.1111/j.1365-2966.2005.09451.x}, \href
  {http://adsabs.harvard.edu/abs/2005MNRAS.363....2K} {363, 2}

\bibitem[\protect\citeauthoryear{{Kormendy} \& {Kennicutt}}{{Kormendy} \&
  {Kennicutt}}{2004}]{Kormendy2004}
{Kormendy} J.,  {Kennicutt} Jr. R.~C.,  2004, \mn@doi [\araa]
  {10.1146/annurev.astro.42.053102.134024}, \href
  {http://adsabs.harvard.edu/abs/2004ARA%26A..42..603K} {42, 603}

\bibitem[\protect\citeauthoryear{{Martig}, {Bournaud}, {Croton}, {Dekel}  \&
  {Teyssier}}{{Martig} et~al.}{2012}]{Martig2012}
{Martig} M.,  {Bournaud} F.,  {Croton} D.~J.,  {Dekel} A.,   {Teyssier} R.,
  2012, \mn@doi [\apj] {10.1088/0004-637X/756/1/26}, \href
  {http://adsabs.harvard.edu/abs/2012ApJ...756...26M} {756, 26}

\bibitem[\protect\citeauthoryear{{Mo}, {Mao}  \& {White}}{{Mo}
  et~al.}{1998}]{Mo1998}
{Mo} H.~J.,  {Mao} S.,   {White} S.~D.~M.,  1998, \mn@doi [\mnras]
  {10.1046/j.1365-8711.1998.01227.x}, \href
  {http://adsabs.harvard.edu/abs/1998MNRAS.295..319M} {295, 319}

\bibitem[\protect\citeauthoryear{{Navarro} \& {Steinmetz}}{{Navarro} \&
  {Steinmetz}}{2000}]{Navarro2000}
{Navarro} J.~F.,  {Steinmetz} M.,  2000, \mn@doi [\apj] {10.1086/309175}, \href
  {http://adsabs.harvard.edu/abs/2000ApJ...538..477N} {538, 477}

\bibitem[\protect\citeauthoryear{{Planck Collaboration} et~al.,}{{Planck
  Collaboration} et~al.}{2014}]{Planck2014}
{Planck Collaboration} et~al., 2014, \mn@doi [\aap]
  {10.1051/0004-6361/201321529}, \href
  {http://adsabs.harvard.edu/abs/2014A%26A...571A...1P} {571, A1}

\bibitem[\protect\citeauthoryear{{Romanowsky} \& {Fall}}{{Romanowsky} \&
  {Fall}}{2012}]{Romanowsky2012}
{Romanowsky} A.~J.,  {Fall} S.~M.,  2012, \mn@doi [\apjs]
  {10.1088/0067-0049/203/2/17}, \href
  {http://adsabs.harvard.edu/abs/2012ApJS..203...17R} {203, 17}

\bibitem[\protect\citeauthoryear{{Rosas-Guevara} et~al.,}{{Rosas-Guevara}
  et~al.}{2015}]{Rosas2015}
{Rosas-Guevara} Y.~M.,  et~al., 2015, \mn@doi [\mnras] {10.1093/mnras/stv2056},
  \href {http://adsabs.harvard.edu/abs/2015MNRAS.454.1038R} {454, 1038}

\bibitem[\protect\citeauthoryear{{Sales}, {Navarro}, {Schaye}, {Dalla Vecchia},
  {Springel}  \& {Booth}}{{Sales} et~al.}{2010}]{Sales2010}
{Sales} L.~V.,  {Navarro} J.~F.,  {Schaye} J.,  {Dalla Vecchia} C.,  {Springel}
  V.,   {Booth} C.~M.,  2010, \mn@doi [\mnras]
  {10.1111/j.1365-2966.2010.17391.x}, \href
  {http://adsabs.harvard.edu/abs/2010MNRAS.409.1541S} {409, 1541}

\bibitem[\protect\citeauthoryear{{Sales}, {Navarro}, {Theuns}, {Schaye},
  {White}, {Frenk}, {Crain}  \& {Dalla Vecchia}}{{Sales}
  et~al.}{2012}]{Sales2012}
{Sales} L.~V.,  {Navarro} J.~F.,  {Theuns} T.,  {Schaye} J.,  {White} S.~D.~M.,
   {Frenk} C.~S.,  {Crain} R.~A.,   {Dalla Vecchia} C.,  2012, \mn@doi [\mnras]
  {10.1111/j.1365-2966.2012.20975.x}, \href
  {http://adsabs.harvard.edu/abs/2012MNRAS.423.1544S} {423, 1544}

\bibitem[\protect\citeauthoryear{{Scannapieco}, {Gadotti}, {Jonsson}  \&
  {White}}{{Scannapieco} et~al.}{2010}]{Scannapieco2010}
{Scannapieco} C.,  {Gadotti} D.~A.,  {Jonsson} P.,   {White} S.~D.~M.,  2010,
  \mn@doi [\mnras] {10.1111/j.1745-3933.2010.00900.x}, \href
  {http://adsabs.harvard.edu/abs/2010MNRAS.407L..41S} {407, L41}

\bibitem[\protect\citeauthoryear{{Schaller}, {Dalla Vecchia}, {Schaye},
  {Bower}, {Theuns}, {Crain}, {Furlong}  \& {McCarthy}}{{Schaller}
  et~al.}{2015}]{Schaller2015}
{Schaller} M.,  {Dalla Vecchia} C.,  {Schaye} J.,  {Bower} R.~G.,  {Theuns} T.,
   {Crain} R.~A.,  {Furlong} M.,   {McCarthy} I.~G.,  2015, \mn@doi [\mnras]
  {10.1093/mnras/stv2169}, \href
  {http://adsabs.harvard.edu/abs/2015MNRAS.454.2277S} {454, 2277}

\bibitem[\protect\citeauthoryear{{Schaye}}{{Schaye}}{2004}]{Schaye2004}
{Schaye} J.,  2004, \mn@doi [\apj] {10.1086/421232}, \href
  {http://adsabs.harvard.edu/abs/2004ApJ...609..667S} {609, 667}

\bibitem[\protect\citeauthoryear{{Schaye} \& {Dalla Vecchia}}{{Schaye} \&
  {Dalla Vecchia}}{2008}]{Schaye2008}
{Schaye} J.,  {Dalla Vecchia} C.,  2008, \mn@doi [\mnras]
  {10.1111/j.1365-2966.2007.12639.x}, \href
  {http://adsabs.harvard.edu/abs/2008MNRAS.383.1210S} {383, 1210}

\bibitem[\protect\citeauthoryear{{Schaye} et~al.,}{{Schaye}
  et~al.}{2015}]{Schaye2015}
{Schaye} J.,  et~al., 2015, \mn@doi [\mnras] {10.1093/mnras/stu2058}, \href
  {http://adsabs.harvard.edu/abs/2015MNRAS.446..521S} {446, 521}

\bibitem[\protect\citeauthoryear{{Seghouane}}{{Seghouane}}{2011}]{AIC2011}
{Seghouane} A.-K.,  2011, \mn@doi [IEEE Transactions on Aerospace Electronic
  Systems] {10.1109/TAES.2011.5751249}, \href
  {http://adsabs.harvard.edu/abs/2011ITAES..47.1154S} {47, 1154}

\bibitem[\protect\citeauthoryear{{Springel}}{{Springel}}{2005}]{Springel2005}
{Springel} V.,  2005, \mn@doi [\mnras] {10.1111/j.1365-2966.2005.09655.x},
  \href {http://adsabs.harvard.edu/abs/2005MNRAS.364.1105S} {364, 1105}

\bibitem[\protect\citeauthoryear{{Springel}, {White}, {Tormen}  \&
  {Kauffmann}}{{Springel} et~al.}{2001}]{Springel2001}
{Springel} V.,  {White} S.~D.~M.,  {Tormen} G.,   {Kauffmann} G.,  2001,
  \mn@doi [\mnras] {10.1046/j.1365-8711.2001.04912.x}, \href
  {http://adsabs.harvard.edu/abs/2001MNRAS.328..726S} {328, 726}

\bibitem[\protect\citeauthoryear{{Stewart}, {Brooks}, {Bullock}, {Maller},
  {Diemand}, {Wadsley}  \& {Moustakas}}{{Stewart} et~al.}{2013}]{Stewart2013}
{Stewart} K.~R.,  {Brooks} A.~M.,  {Bullock} J.~S.,  {Maller} A.~H.,  {Diemand}
  J.,  {Wadsley} J.,   {Moustakas} L.~A.,  2013, \mn@doi [\apj]
  {10.1088/0004-637X/769/1/74}, \href
  {http://adsabs.harvard.edu/abs/2013ApJ...769...74S} {769, 74}

\bibitem[\protect\citeauthoryear{{Thacker} \& {Couchman}}{{Thacker} \&
  {Couchman}}{2001}]{Thacker2001}
{Thacker} R.~J.,  {Couchman} H.~M.~P.,  2001, \mn@doi [\apjl] {10.1086/321739},
  \href {http://adsabs.harvard.edu/abs/2001ApJ...555L..17T} {555, L17}

\bibitem[\protect\citeauthoryear{{Toomre}}{{Toomre}}{1977}]{Toomre1977}
{Toomre} A.,  1977, in {Tinsley} B.~M.,  {Larson} D.~Campbell R.~B.~G.,  eds,
  Evolution of Galaxies and Stellar Populations. p.~401

\bibitem[\protect\citeauthoryear{{Velliscig} et~al.,}{{Velliscig}
  et~al.}{2015}]{Vell2015}
{Velliscig} M.,  et~al., 2015, \mn@doi [\mnras] {10.1093/mnras/stv1690}, \href
  {http://adsabs.harvard.edu/abs/2015MNRAS.453..721V} {453, 721}

\bibitem[\protect\citeauthoryear{{Vogelsberger} et~al.,}{{Vogelsberger}
  et~al.}{2014}]{Vogelsberger2014}
{Vogelsberger} M.,  et~al., 2014, \mn@doi [\mnras] {10.1093/mnras/stu1536},
  \href {http://adsabs.harvard.edu/abs/2014MNRAS.444.1518V} {444, 1518}

\bibitem[\protect\citeauthoryear{{Weil}, {Eke}  \& {Efstathiou}}{{Weil}
  et~al.}{1998}]{Weil1998}
{Weil} M.~L.,  {Eke} V.~R.,   {Efstathiou} G.,  1998, \mn@doi [\mnras]
  {10.1046/j.1365-8711.1998.01931.x}, \href
  {http://adsabs.harvard.edu/abs/1998MNRAS.300..773W} {300, 773}

\bibitem[\protect\citeauthoryear{{White}}{{White}}{1984}]{White1984}
{White} S.~D.~M.,  1984, \mn@doi [\apj] {10.1086/162573}, \href
  {http://adsabs.harvard.edu/abs/1984ApJ...286...38W} {286, 38}

\bibitem[\protect\citeauthoryear{{White} \& {Frenk}}{{White} \&
  {Frenk}}{1991}]{White1991}
{White} S.~D.~M.,  {Frenk} C.~S.,  1991, \mn@doi [\apj] {10.1086/170483}, \href
  {http://adsabs.harvard.edu/abs/1991ApJ...379...52W} {379, 52}

\bibitem[\protect\citeauthoryear{{White} \& {Rees}}{{White} \&
  {Rees}}{1978}]{White1978}
{White} S.~D.~M.,  {Rees} M.~J.,  1978, \mnras, \href
  {http://adsabs.harvard.edu/abs/1978MNRAS.183..341W} {183, 341}

\bibitem[\protect\citeauthoryear{{Wiersma}, {Schaye}  \& {Smith}}{{Wiersma}
  et~al.}{2009a}]{Wiersma2009a}
{Wiersma} R.~P.~C.,  {Schaye} J.,   {Smith} B.~D.,  2009a, \mn@doi [\mnras]
  {10.1111/j.1365-2966.2008.14191.x}, \href
  {http://adsabs.harvard.edu/abs/2009MNRAS.393...99W} {393, 99}

\bibitem[\protect\citeauthoryear{{Wiersma}, {Schaye}, {Theuns}, {Dalla Vecchia}
   \& {Tornatore}}{{Wiersma} et~al.}{2009b}]{Wiersma2009b}
{Wiersma} R.~P.~C.,  {Schaye} J.,  {Theuns} T.,  {Dalla Vecchia} C.,
  {Tornatore} L.,  2009b, \mn@doi [\mnras] {10.1111/j.1365-2966.2009.15331.x},
  \href {http://adsabs.harvard.edu/abs/2009MNRAS.399..574W} {399, 574}

\bibitem[\protect\citeauthoryear{{Zavala}, {Okamoto}  \& {Frenk}}{{Zavala}
  et~al.}{2008}]{Zavala2008}
{Zavala} J.,  {Okamoto} T.,   {Frenk} C.~S.,  2008, \mn@doi [\mnras]
  {10.1111/j.1365-2966.2008.13243.x}, \href
  {http://adsabs.harvard.edu/abs/2008MNRAS.387..364Z} {387, 364}

\bibitem[\protect\citeauthoryear{{Zavala}, {Avila-Reese}, {Firmani}  \&
  {Boylan-Kolchin}}{{Zavala} et~al.}{2012}]{Zavala2012}
{Zavala} J.,  {Avila-Reese} V.,  {Firmani} C.,   {Boylan-Kolchin} M.,  2012,
  \mn@doi [\mnras] {10.1111/j.1365-2966.2012.22100.x}, \href
  {http://adsabs.harvard.edu/abs/2012MNRAS.427.1503Z} {427, 1503}

\makeatother
\end{thebibliography}

\bsp	
\label{lastpage}

\end{document}